\documentclass[twocolumn,english,aps,superscriptaddress,prb,showpacs]{revtex4-1}
\usepackage{amsmath}
\usepackage{amssymb}
\usepackage{graphicx}
\usepackage[colorlinks,bookmarks=false,citecolor=red,linkcolor=blue,urlcolor=blue]{hyperref}
\def\Hhat{\hat{H}}

\def\Sbf{\textbf{S}}

\def\i{{\boldsymbol i}}
\def\j{{\boldsymbol j}}

\def\S{{\boldsymbol S}}
\def\mibf{{\mathbf{{\it m}}}}
\def\mjbf{{\mathbf{{\it m^\prime}}}}
\def\mbf{{\mathbf{{\it m}_1}}}
\def\mmbf{{\mathbf {{\it m}_2}}}

\def\identity{\hat{\mathbf{1}}}
\newcommand{\ve}[1]{\boldsymbol{#1}}
\begin{document}
\title{Exploring the Kondo effect of an extended impurity with chains of Co adatoms in a magnetic field}
\author{Bimla Danu}
\email{danu.bimla@epfl.ch}
\affiliation{Institute of Physics, \'Ecole Polytechnique F\'ed\'erale de Lausanne (EPFL), CH-1015 Lausanne, Switzerland}
\author{Fakher Assaad}
\email{assaad@physik.uni-wuerzburg.de}
\affiliation{Institut f\"ur Theoretische Physik und Astrophysik, Universit\"at W\"urzburg, Am Hubland,
D-97074 W\"urzburg, Germany} 
\affiliation{W\"urzburg-Dresden Cluster of Excellence ct.qmat }
\author{Fr\'ed\'eric Mila}
\email{frederic.mila@epfl.ch}
\affiliation{Institute of Physics, \'Ecole Polytechnique F\'ed\'erale de Lausanne (EPFL), CH-1015 Lausanne, Switzerland}
\date{\today}
\begin{abstract}
Motivated by recent STM experiments, we  explore the magnetic field induced Kondo effect that takes place at symmetry protected level crossings in finite Co adatom  chains.  We argue that the effective two-level system realized at a level crossing acts as an extended impurity coupled to the conduction electrons of the substrate by a distribution of Kondo couplings at the sites of the chain. Using auxiliary-field quantum Monte Carlo simulations, which quantitatively reproduce the field dependence of the zero-bias signal, we show that a proper Kondo resonance is present at the sites where the effective Kondo coupling dominates. Our modeling and  numerical simulations  provide a theoretical basis for the interpretation of the STM spectrum in terms of level crossings of the Co adatom chains. 
\end{abstract}
\maketitle
The Kondo effect is one of the most extensively studied and adequately addressed many-body process occurring due to the screening of a local moment by a conduction electron cloud~\cite{hewson1993,Kondo1964,Wilson1975}. On the experimental side,  recent advances in   scanning tunneling microscopy (STM)  open greater opportunities to realise and investigate the Kondo effect in various Kondo nanostructures~\cite{Madhavan1998,Cronenwett1999,Park2002,Herrero2005,Wahl2005,Neel2011,Spinelli2015,Toskovic2016}.  For instance, the recent  STM  experiments on finite atomic spin-chain realizations of Co adatoms in the presence of an external magnetic field have revealed an interesting interplay between the Kondo problem and the physics of quantum spin chains in a field~\cite{Toskovic2016}. Co adatoms on a Cu$_2$N/Cu(100) surface carry a spin-3/2 with a strong uniaxial hard-axis anisotropy($D$)~\cite{Otte2008,Spinelli2015}, and applying an external magnetic field perpendicular to the surface, the Co adatom chain effectively behaves like a spin-1/2 XXZ chain in transverse field.
The magnetic field induced level crossings of  finite XXZ  and SU(2) chains are similar so that the experiments of  Ref.~[\onlinecite{Toskovic2016}]  can be discussed in the context of the  SU(2) invariant version of the model:
\begin{eqnarray}
\Hhat & = & -t\sum_{\langle \i,\j\rangle, \sigma}(\hat c^{\dagger}_{\i,\sigma} \hat c_{\j,\sigma}+h.c)+J_k \sum^L_{\ve{l} =1} \hat{\ve{S}}^c_{\ve{l}} \cdot \hat{\ve{S}}_{\ve{l}}  \nonumber \\ 
      & & +J_h\sum^{L-1}_{\ve{l}=1} \hat{\ve{S}}_{\ve{l}} \cdot \hat{\ve{S}}_{\ve{l}+\Delta\ve{l}} -g \mu_B h^{\bf z} \sum^{L}_{\ve{l}=1} \hat{S}^z_{\ve{l}}.
\label{model_ham}
\end{eqnarray}
Here, $t$ is the hopping parameter of the conduction electrons, $J_k$ the antiferromagnetic Kondo coupling between a Co adatom and the conduction electrons, $J_h$  the Heisenberg antiferromagnetic coupling, $h^{\bf z}$  an external magnetic field in the $z$ direction, $L$ the length of the Heisenberg chain, $\hat{\ve{S}}_{\ve{l}}$ spin-1/2 operators and $\hat{\ve{S}}^c_{\ve{l}}=\frac{1}{2}\sum_{\sigma,\sigma^\prime} \hat{c}^{\dagger}_{\ve{l}, \sigma} \ve{\sigma}_{\sigma,\sigma^\prime} \hat{c}^{}_{\ve{l},\sigma^\prime}$ denotes the spin of conduction electrons. Throughout the calculation we set $t,\mu_B=1$ and $g=2$. 

 When  the Kondo coupling is switched off ($J_k=0$), the chain undergoes a series of level crossings that lead to steps in the magnetization curve. In the geometry used in the experiment of 
 Ref.~[\onlinecite{Toskovic2016}], $J_h$  and the Kondo energy $\epsilon_k=k_B T_k$ are both of the order of 0.2 meV, with two important consequences: There is a competition between the Heisenberg coupling and the Kondo effect, and one can reach the saturation field of the isolated chain.

The main result of the STM experiments of Ref.~[\onlinecite{Toskovic2016}] is to demonstrate that the differential conductance exhibits a series of anomalies as a function of the field, and that these anomalies coincide with the fields at which the isolated chain is expected to undergo level crossings. These anomalies are strongly site dependent however. Occasionally they take the typical V-shape of the Kondo resonance of a single impurity split by a magnetic field, but in most cases they are less pronounced if at all. 

In this Letter, our goal is to provide a theory of STM that is valid at low temperature and that puts the measurements in the appropriate Kondo context. As we shall see, a Kondo effect is indeed present at each level crossing,  but it corresponds to that of an extended impurity of size the length of the chain.   The Hilbert space of this extended impurity corresponds to the twofold degenerate ground state of the spin chain at the level crossing.  The presence or absence of a Kondo resonance at a given Co  site where the STM signal is recorded  depends i)  on a matrix element encoding the fact that the ground state of the spin chain is addressed at a given Co site,  and ii)  on the magnitude of the effective Kondo coupling between the extended impurity and the substrate at the considered Co location. 

\noindent{\textit{Method.}}
At high temperatures, the differential conductance in the presence of the Kondo coupling can be calculated using perturbation theory.   While the results reproduce the gross features of the experimental data,  they are  limited to  a regime above the Kondo temperature, and  cannot account for the details of the low temperature data of Ref.~[\onlinecite{Toskovic2016}].  
For a particle-hole symmetric conduction band,   our model can be simulated with the  auxiliary field quantum Monte Carlo   (QMC)   algorithm  without encountering the negative sign problem.   We have  used the finite temperature  algorithm  \cite{Blankenbecler81,White89,Assaad08_rev}   of the ALF-project \cite{ALF_v1}  and followed   Refs.~[\onlinecite{Assaad99a,SatoT17_1}] for the implementation of our Kondo model.   In the QMC calculation  we consider a $20\times20$ square lattice   with unit lattice constant  and hopping matrix element $t$ and  consider  a linear arrangement of  magnetic adatoms  at distance   $\Delta \ve{l} =(0,3)$ or $ \Delta \ve{l} = (3,2) $    from each other (up to $L=7$). To overcome the finite size effects  we included an orbital magnetic field corresponding to a single flux quantum traversing the whole lattice \cite{Assaad2002}, and a rather large value of the Kondo  interaction   $J_k/t=2$  so as to  ensure that the Kondo  scale of the single impurity problem remains larger than the finite size level spacing of the conduction electrons. Finally we  consider $J_h/t=1.8$.

In the case of a single adatom ($L=1$), the problem reduces to that of a single impurity.   The low temperature STM signal observed in Ref.~[\onlinecite{Toskovic2016}] consists of a single peak, the  Kondo resonance,  consistent with a tunneling process from sample to tip that goes through the localized d-orbital of the  Co adatoms.  To  account for this in the realm of the Kondo model  we compute  co-tunneling processes \cite{Figgins2010,Ternes2015,Morr2017}  given by: $ A_{\ve{l}}(\omega)    =   - \text{Im}  G_{\ve{l}}^{\text{ret}} (\omega ) $ with  $ G_{\ve{l}}^{\text{ret}}  (\omega)  =   - i \int_{0}^{\infty} d\tau e^{i \omega \tau}  \sum_{\sigma } \big< \big\{ \tilde{d}_{\ve{l},\sigma}^{}(\tau), \tilde{d}_{\ve{l},\sigma}^{\dagger} (0) \big\} \big>  $  and  $ \tilde{d}_{\ve{l},\sigma}^{\dagger} =  \hat{c}^{\dagger}_{\ve{l}, -\sigma} \hat{S}^{\sigma}_{\ve{l}} + \sigma  \hat{c}_{\ve{l},\sigma}^{\dagger} \hat{S}^{z}_{\ve{l}}$.  Here $\sigma = \pm$   runs over the two spin polarization and $   \hat{S}^{\pm}_{\ve{l}} = \hat{S}^{x}_{\ve{l}}  \pm i \hat{S}^{y}_{\ve{l}} $.   This form can be obtained  by starting from the  single impurity Anderson model  and applying  the canonical  Schrieffer-Wolf  transformation (see supplemental material of Ref.~[\onlinecite{Raczkowski18}])   and  agrees  with  the  expression given in Ref.~[\onlinecite{TCosti2000}].    

\begin{figure}[htbp]
\centering
\includegraphics[width=0.47\textwidth]{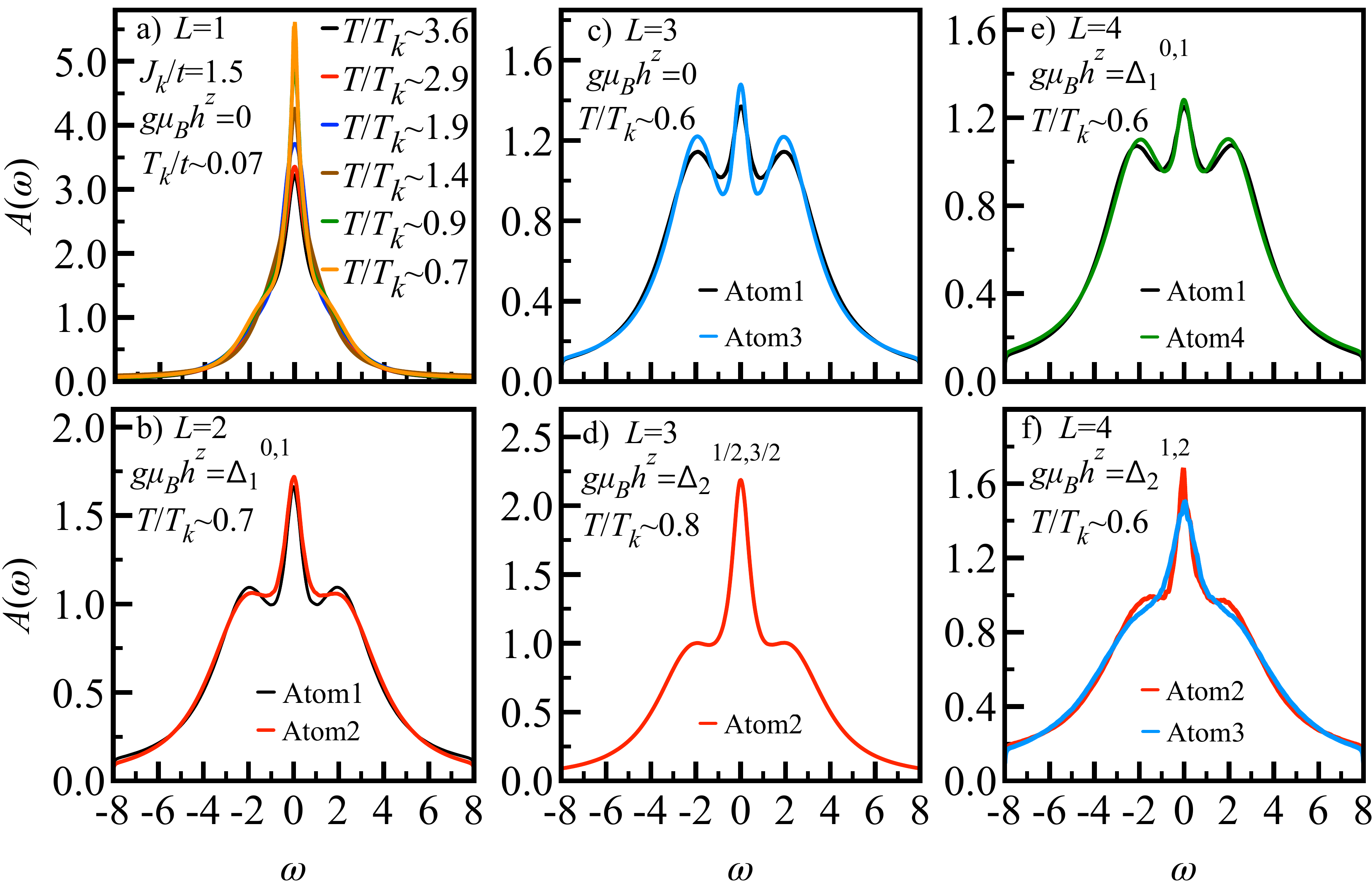}
\caption{The spectral function computed using  stochastic analytical continuation algorithm~\cite{KBeach2004} at a given level crossings up to $L=4$.  For $L>1$ we choose $J_h/t=1.8$ and $J_k/t=2$. The corresponding Kondo scale is extracted in Fig.~\ref{fig:Tchi_vs_T_L1234}.}
\label{fig:Aomega_vs_omega_L1234}
\end{figure}

Let us start by showing examples of spectral functions at level crossings obtained from QMC by stochastic analytic continuation~\cite{KBeach2004}.
As apparent from  Fig.~\ref{fig:Aomega_vs_omega_L1234}a),  for a single impurity we observe the characteristic  temperature dependence of a Kondo resonance at zero field. 
Fig.~\ref{fig:Aomega_vs_omega_L1234}  also shows the magnetic field induced Kondo resonances. For the two site chain there is a single level crossing  between the singlet 
and  the triplet   
at $g\mu_Bh^z=\Delta_1^{0,1}$.    In the generic Kondo problem, time reversal symmetry protects the two-fold degeneracy of the 
impurity state.     Here   parity  protects the level crossings and  a Kondo  resonance is apparent on  both adatom sites, see Fig.~\ref{fig:Aomega_vs_omega_L1234}b).  For the three site  chain  two level crossings occur before saturation. The ground state  is a spin-1/2 doublet in zero field and  resonances are seen on the first and third adatoms, see Fig.~\ref{fig:Aomega_vs_omega_L1234}c).    At the second level crossing, the resonance is seen only on  the central adatom, see Fig.~\ref{fig:Aomega_vs_omega_L1234}d). For $L=4$, Kondo resonances emerge on outer adatoms at the first level crossing, see Fig.\ref{fig:Aomega_vs_omega_L1234}e), and on the central adatoms for the second level crossing, see Fig.~\ref{fig:Aomega_vs_omega_L1234}.f).

These results have been obtained at temperatures already representative of the low temperature regime, and they reproduce the main features of the experimental results (see Supplemental Material, Ref.~[\onlinecite{suppl}], Fig.~\ref{fig:dIbydV_vs_V_STM}). However, to make a quantitative comparison with the experiments, which correspond to much lower temperature, 
we will concentrate on the zero bias differential conductance measured in the 
STM experiment~\cite {Spinelli2015,Toskovic2016,Otte2008} as
 \begin{equation}
 \begin{aligned}
dI_{\ve{l}}/dV(V=0) =2\frac{e^2}{\hbar} \int^{\infty}_{-\infty} d \omega \Big(-\frac{df(\omega)}{d\omega}\Big)  A_{\ve{l}}(\omega)
\end{aligned}
\end{equation}
where $f(\omega)$ is a Fermi function.  In the low temperature limit the above maps onto: 
 \begin{equation}
dI_{\ve{l}}/dV(V=0)   \simeq 2\frac{e^2}{\hbar} A_{\ve{l}}(\omega=0)  \simeq 2\frac{e^2}{\pi\hbar}   \beta G_{\ve{l}}(\tau = \beta/2)
 \end{equation}
 where $G_{\ve{l}} (\tau) = \sum_{\sigma}\langle \tilde{d}_{\ve{l},\sigma}^{}(\tau) \tilde{d}_{\ve{l},\sigma}^{\dagger}(0)   \rangle $ is the imaginary time Green function which can be directly computed in the auxiliary field QMC.   This approach avoids analytical continuation and  our discussion will be based on the field dependence of this quantity.    In the zero temperature limit the above equation is exact,  and a more precise account of the zero bias differential conductance at finite temperature  without  using  the analytical continuation  can be obtained following Refs.~[\onlinecite{Luitz11,Karrasch10,Monien10}].
 \begin{figure}
\includegraphics[width=0.45\textwidth]{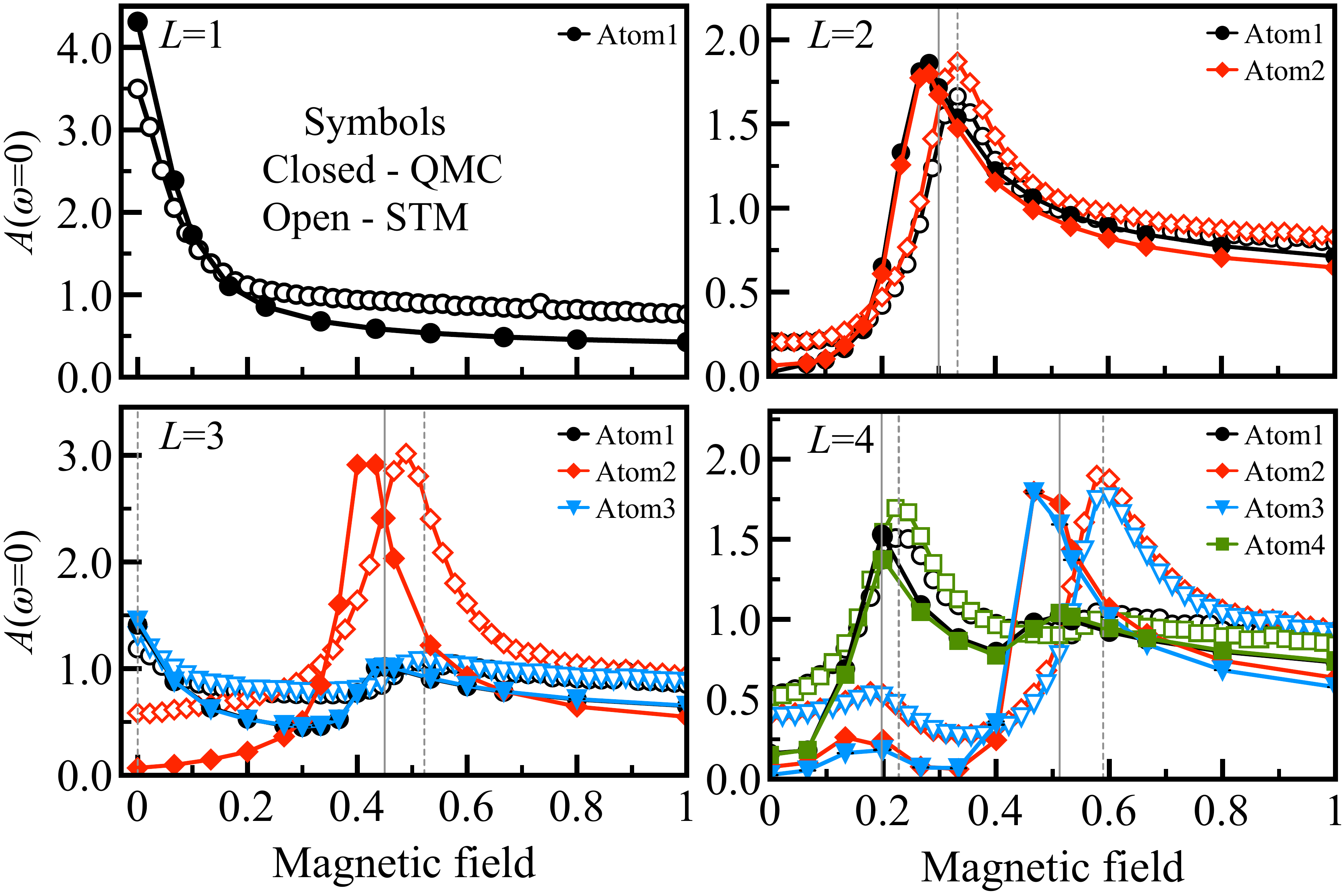}
\caption{The $\tilde d$-spectral function at $\omega=0$ computed by QMC for a Heisenberg chain in a field together with the zero bias conductance measured in the STM experiment (in atomic units) for an XXZ chain in transverse field~\cite{Toskovic2016}. The magnetic field axis is normalised by the maximum values in both cases. The continuous grey vertical lines and the dashed grey lines denote the expected exact positions of the level crossings for a Heisenberg chain and for an XXZ one~\cite{Toskovic2016}, respectively.}
 \label{comp_Greenf}
\end{figure}

The QMC results of the  local spectral function at zero frequency for $k_BT/t=1/30$ are compared to the zero bias conductance reported in Ref.~[\onlinecite{Toskovic2016}] as a function of  external magnetic field in Fig.~\ref{comp_Greenf}. Noticeably, up to four atoms the zero frequency spectral function shows excellent agreement with the corresponding zero bias conductance measured in the experiment.
 The temperature scales   in the   QMC and  STM  are comparable: the data presented in Fig.~\ref{comp_Greenf} are  computed below $k_BT^l_k/8t$, where $T^l_k$ is an estimate of Kondo temperature from scaling of local spin susceptibility~\cite{Hirsch1986,Assaad2002} at each level crossings (see below), while the STM  data are taken at $330$ mK $\sim T^{Co}_k/8$ ($k_BT/t\sim1/35$ in QMC). 
 
\begin{figure}[htbp]
\centering
\includegraphics[width=0.45\textwidth] {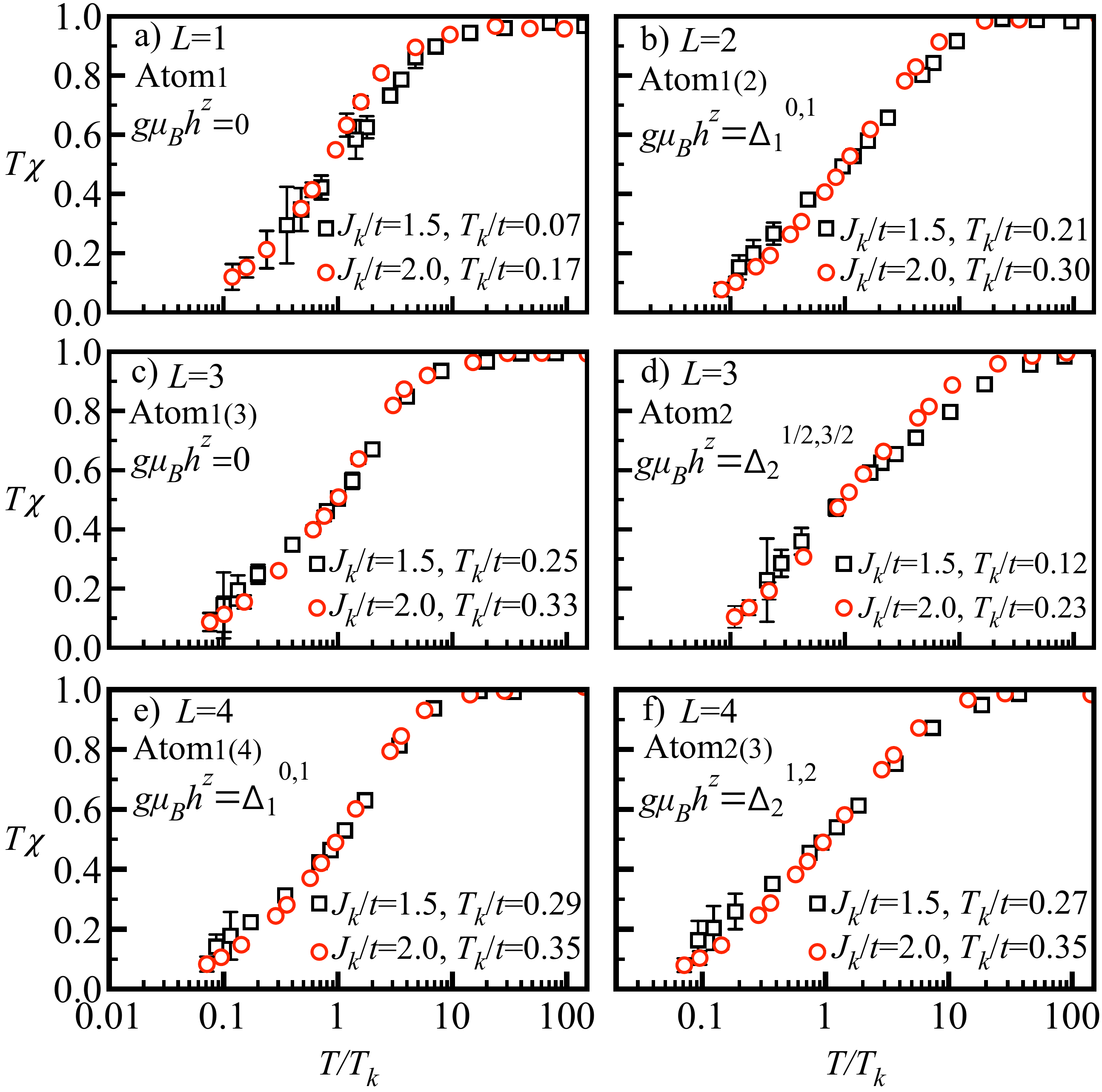}
\caption{Local probe transverse susceptibility as a function of temperature  computed for $J_h/t=1.8$  up to $L=4$  at sites where the level crossing leads to a Kondo resonance. The sites not shown do not follow the typical temperature dependence on Kondo couplings up to $k_BT/t\approx1/40$.} 
\label{fig:Tchi_vs_T_L1234}
\end{figure}

\begin{figure}[htbp]
\centering
\includegraphics[width=0.45\textwidth]{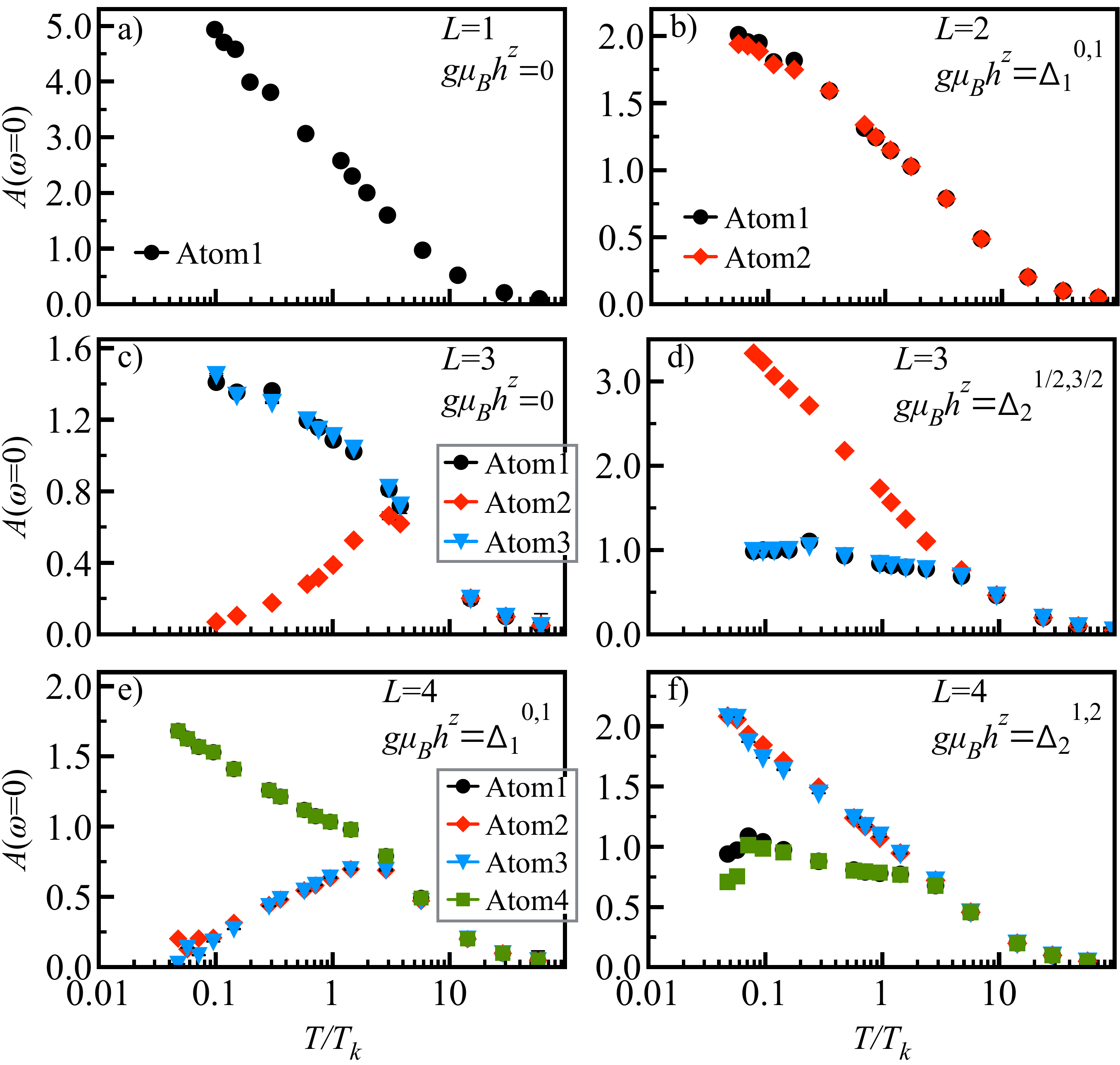}%
\caption{Exact zero-bias conductance as a function of temperature normalised by the corresponding Kondo scale at each level crossings up to $L=4$.} 
\label{fig:peakheight_vs_T}
\end{figure}

 To associate an adatom dependent Kondo temperature to each level crossing, we compute the  local transverse susceptibility, $\chi_{\ve{l}}=\int^{\beta}_{0} d \tau\langle \hat S^{+}_{\ve{l}}(\tau)\hat S^{-}_{\ve{l}}(0)+h.c \rangle$, where $\hat S^{\pm}_{\ve{l}}(\tau)=e^{\tau \hat H} \hat S^\pm_{\ve{l}} e^{-\tau \hat H}$.   Interestingly we  observe  in Fig.~\ref{fig:Tchi_vs_T_L1234}  that when the STM  data at an adatom site   shows a resonance, the local susceptiblity follows the  expected universal behaviour,  $T\chi_l  =   f(T/T^l_k)$, where  $T^l_k$   corresponds to the    adatom   and level crossing resolved Kondo temperature.   
Fig.~\ref{fig:peakheight_vs_T}  plots the temperature dependence of the zero frequency spectral function as estimated by $A_{\ve{l}}(\omega=0) \simeq 
\frac{1}{\pi}  \beta G_{\ve{l}}(\tau = \beta/2) $.    For $L=3$ and $L=4$ strong site dependence of the signal emerges below the Kondo temperature.  We first concentrate on cases where we observe a dominant resonance. For these cases, the temperature dependence of the zero-bias conductance at the various sites (see Fig.~\ref{fig:peakheight_vs_T})  shows logarithmic increase at the Kondo scale  upon reducing the temperature.     At the other sites no  such increase is observed.  Quite remarkably, depending on the site, level crossings can show up as a peak, a dip, or a change of slope as a function of field. Accordingly, in the experimental results of Ref.~[\onlinecite{Toskovic2016}], the frequency dependence at a critical field may or may not show a Kondo resonance.

\textit{Effective model.}  To provide an effective model for a given level crossing, $p$,  we project the Hamiltonian  on the two fold  degenerate Hilbert space of the   level crossing.  Clearly   such a  strategy is valid only  if the gap separating the  next excited states of the  spin chain is large  compared to the effective Kondo scale.  For our SU(2) model, this approximation will necessarily fail in the large $L$ limit,  but as we will see below it provides an accurate account of the QMC data for small $L$.    Let $\{|\mbf\rangle=|\Sbf_\mbf\Sbf^z_\mbf\rangle_p,|\mmbf \rangle = | \Sbf_\mmbf\Sbf^z_\mmbf\rangle_p\}$  be the eigenstate of the Heisenberg chain  with   energies  $e_{m_1,p} $ , $e_{m_2,p} $ that span the Hilbert space of the level crossing and  $\hat{P} $ the projector onto this space.     Defining a vector of Pauli spin matrices $\ve{\tau}$ that act on this Hilbert space, the projected Hamiltonian, up to a constant, reads: 
\begin{eqnarray}
&&\Hhat^{eff}_p=-t\sum_{\langle \i,\j\rangle, \sigma}(\hat c^{\dagger}_{\i,\sigma} \hat c_{\j,\sigma}+h.c) +    \nonumber
J_k \sum_{\ve{l}}\Big\{j^{\perp}_{\ve{l},p}(\hat S^{x,c}_{\ve{l}} \hat{\tau}^x+ \\ &&\hat{S}^{y,c}_{\ve{l}}  \hat \tau^y)+j^{z}_{\ve{l},p}\hat S^{z,c}_{\ve{l}} \hat{\tau}^z+\mu_{\ve{l},p} S^{z,c}_{\ve{l}}+(\Delta^{\mbf,\mmbf}_p-g\mu_Bh^{\bf z})\hat n\Big\}  
\label{effe_ham}
\end{eqnarray}
The energy difference between the two local states  is given by $\Delta^{\mbf,\mmbf}_p=e_{\mmbf,p}-e_{\mbf,p}$,  $\hat{n}=\frac{1}{2}(\identity+ \hat {\tau}^z)$, and the site dependent effective couplings and the effective magnetic field are defined as;
  $j^{\perp}_{\ve{l},p}=\langle \mmbf|\hat{S}^{x,y}_{\ve{l}} |\mbf\rangle$, $2j^{z}_{\ve{l},p}=-\langle \mbf|\hat{S}^z_{\ve{l}} |\mbf \rangle+\langle \mmbf|\hat{S}^z_{\ve{l}} |\mmbf \rangle$  and $2\mu_{\ve{l},p}=\langle \mbf|\hat{S}^z_{\ve{l}} |\mbf \rangle+\langle \mmbf| \hat{S}^z_{\ve{l}} |\mmbf \rangle$.
 \begin{figure}
\includegraphics[width=0.48\textwidth]{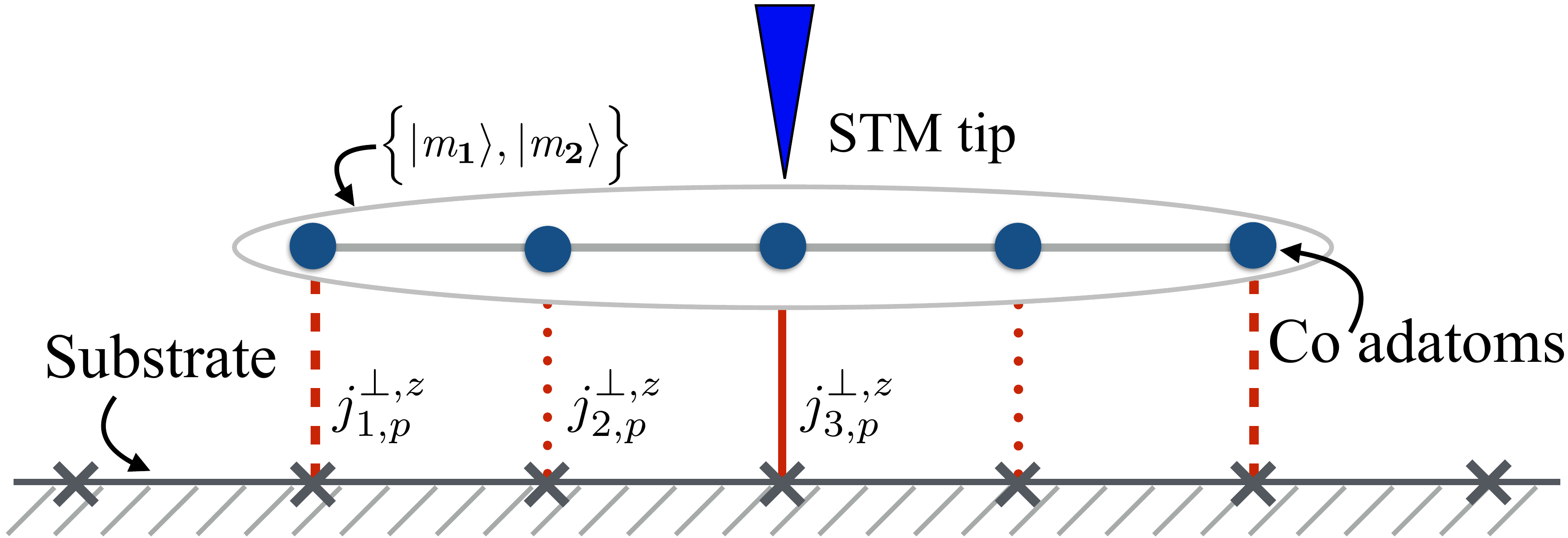}
\caption{A chain of Co adatoms on a substrate implement, at a level crossing, the Kondo model of an extended impurity.}
\label{Coadatoms_skech}
\end{figure}
\begin{table}
\caption{\label{tab:j_x,z}Effective couplings $j^{\perp,z}_{\ve{l}(\ve{l}^\prime),p}$ at the level crossings($p=1, \cdots,\frac{L}{2}(\frac{L+1}{2})$ for a fixed even(odd) $L$.)}
\begin{ruledtabular}
\begin{tabular}{l | l}
$L=2$& $j^{\perp}_{1(2),1}= \frac{1}{2\sqrt{2}}$, \quad $j^{z}_{1(2),1}=\frac{1}{4}$\\
$L=3$& $j^{\perp}_{1(3),1}=\frac{1}{3}$,\quad $j^{\perp}_{2,1}=\frac{1}{6}$, \quad $j^{z}_{1(3),1}=\frac{1}{3}$, \quad   $j^{z}_{2,1}=\frac{1}{6}$\\
$L=3$& $j^{\perp}_{1(3),2}=\frac{1}{2\sqrt{6}}$, \quad  $j^{\perp}_{2,2}=\frac{1}{\sqrt{6}}$, \quad $j^{z}_{1(3),2}=\frac{1}{12}$,\quad $j^{z}_{2,2}=\frac{1}{3}$\\
$L=4$ & $j^{\perp}_{1(4),1}=\frac {1}{4 \sqrt{6}{\sqrt{2+\sqrt{2}}}}\Big[ \frac{1+\sqrt{2}}{1+\sqrt{3}}+1+\sqrt{1+\frac{\sqrt{3}}{2}}\big(1+\sqrt{2}\big)\Big]$\\
& $j^{\perp}_{2(3),1}= \frac {1}{4 \sqrt{6}{\sqrt{2+\sqrt{2}}}}\Big[ \frac{1}{1+\sqrt{3}}+(1+\sqrt{2})+\sqrt{1+\frac{\sqrt{3}}{2}}\Big]$\\
&  $j^{z}_{1(4),1}=\frac{1}{16} (2+\sqrt{2})$, \quad $j^{z}_{2(3),1}=\frac{1}{16} (2-\sqrt{2})$\\
$L=4$& $j^{\perp}_{1(4),2}=\frac {1}{4{\sqrt{2+\sqrt{2}}}} $, \quad  $j^{\perp}_{2(3),2}= \frac {1+\sqrt{2}}{4{\sqrt{2+\sqrt{2}}}}$\\
& $j^{z}_{1(4),2}=\frac{1}{16} (2-\sqrt{2})$, \quad $j^{z}_{2(3),2}=\frac{1}{16} (2+\sqrt{2})$\\
\end{tabular}
\end{ruledtabular}
\end{table}

This model can be interpreted as the Kondo model of an extended impurity that is coupled to the conduction electrons at $L$ points (see Fig.~\ref{Coadatoms_skech}).   The projection does not affect the U(1) spin symmetry of the model, but  as shown in Table~\ref{tab:j_x,z} it yields a strong site  dependence of the effective couplings  $j^{\perp,z}_{\ve{l},p}$.     To compute the  co-tunneling  within the effective model, we still have to project the spin operator  onto the level-crossing  Hilbert space 
$ \hat{P} \tilde{d}_{\ve{l},\sigma}^{\dagger} \hat{P} =  \hat{c}^{\dagger}_{\ve{l}, -\sigma}   \hat{P}  \hat{S}^{\sigma}_{\ve{l}}  \hat{P} + \sigma  \hat{c}_{\ve{l},\sigma}^{\dagger}  \hat{P} S^{z}_{\ve{l}}  \hat{P} $ so that it acquires a site dependence when written in terms of  $\ve{\tau}$ operators. This reflects  the fact that in  the experiment  the extended states  $| \mbf \rangle $ and $ | \mmbf \rangle $ are addressed  via manipulation of  one of the constituent Co d-spins.   A detailed numerical analysis of  the effective model along these lines is left for future studies.  However, it is already clear that  it provides  a qualitative interpretation of the data in terms of a projection induced hierarchy of Kondo scales (see Ref.~[\onlinecite{suppl}], Table~\ref{tab:epsilonk_L1234567}). 
Consider for instance the four-site chain at the first, $p=1$, level crossing  for which 
$j^{z}_{1(4),1} / j^{z}_{2(3),1} = 5.83$ and  $j^{\perp}_{1(4),1}/j^{\perp}_{2(3),1}  = 1.25 $.  Thereby, the  Kondo singlet will be predominantly formed by  an entangled state of the degenerate two level system, $ \left\{ | \mbf \rangle, | \mmbf \rangle  \right\}  $  and a symmetric combination of the conduction electron spins on sites  one and four (see Ref.~[\onlinecite{suppl}], Section~\ref{Effec_Kondo_scale}.).  This provides  an understanding of the observed Kondo like  temperature dependence of the zero-bias conductance at sites one and four.   

\begin{figure}
\includegraphics[width=0.45\textwidth]{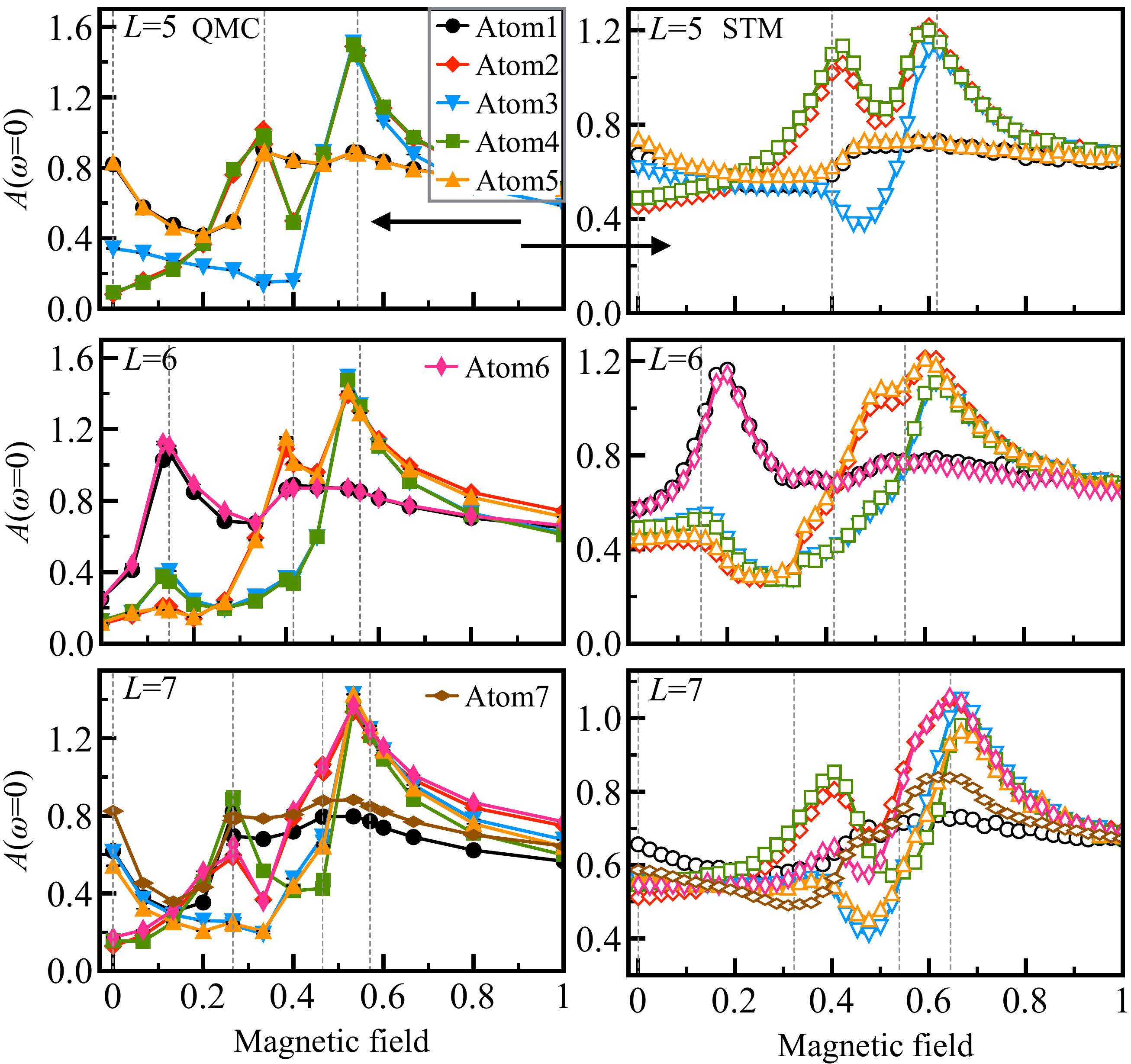}
\caption{QMC results for five, six and seven atom spin-chains together with the STM data around $V\sim0.6$ mV~\cite {Toskovic2016} as a function of magnetic field (normalised by the maximum values).}
\label{comp_567}
\end{figure}

In Fig.~\ref{comp_567} we consider   chains up to seven spins. Here, an accurate comparison of the zero bias conductance  is hard due to the asymmetric line shape of the Kondo resonances that arise in the experiment due to the potential scattering between tip and sample~\cite{Figgins2010,von2015,Ternes2015,Morr2017}, but a reasonable agreement can be achieved slightly away from the zero bias around $V\sim0.4-0.8$ mV as shown in Fig.~\ref{comp_567}.  
As the chain size grows,  the spectrum  will collapse, and our understanding in terms of the projection onto the two-fold level crossing Hilbert space fails. In fact in this limit  we expect a  crossover to Kondo lattice behavior  characterized at the mean field level  by hybridized heavy and light bands in a magnetic field (see Ref.~[\onlinecite{suppl}],  Section~\ref{LargeN_meanfield}).  For an infinite  Heisenberg chain  and within this approximation, we expect the  conductance to reflect the local  spinon density of states.     

To summarize, we have shown that the STM measurements on Co adatom chains agree remarkably well with QMC simulations of a model of spin-1/2 chains in a field coupled to a conducting substrate via local Kondo couplings.  To interpret the strong site dependence of the signal, we have performed a projection onto the symmetry protected level crossing  Hilbert  space  of the spin chain that leads to the notion of  an extended  impurity  with  site dependent  Kondo  couplings.  Consequently,   screening happens via the dominant channel such that  site  and level crossing  dependent  Kondo  resonances are  observed in the  STM  co-tunneling   conductance as well as in the Monte Carlo simulations.  As a function of chain length, the above construction will  progressively fail and we expect a crossover to   Kondo lattice physics, in which the STM  signal will pick up the two spinon continuum of the   spin chain.  Further work at understanding  the details of the extended impurity Kondo model, as well as the crossover to the lattice limit is presently under  progress. 
  
\begin{acknowledgments}
We thank  S. Otte, M. Raczkowski, and M. Ternes for very useful discussions and S. Otte and M. Ternes for providing the STM data. FFA thanks  the DFG collaborative research centre SFB1170 To CoTronics (project C01)   for financial support as well as the W\"urzburg-Dresden Cluster of Excellence on Complexity and Topology in Quantum Matter -- \textit{ct.qmat} (EXC 2147, project-id 39085490). The authors gratefully acknowledge the Gauss Centre for Supercomputing e.V. (www.gauss-centre.eu) for funding this project by providing computing time on the GCS Supercomputer SUPERMUC-NG at Leibniz Supercomputing Centre (www.lrz.de). FM and BD acknowledge the Swiss National Science Foundation t and its SINERGIA network “Mott physics beyond the Heisenberg model” for financial support.
This work was also supported by EPFL through the use of the facilities of its Scientific IT and Application Support Center.
\end{acknowledgments}
\bibliography{Kondo_ref}

\begin{thebibliography}{36}%
\makeatletter
\providecommand \@ifxundefined [1]{%
 \@ifx{#1\undefined}
}%
\providecommand \@ifnum [1]{%
 \ifnum #1\expandafter \@firstoftwo
 \else \expandafter \@secondoftwo
 \fi
}%
\providecommand \@ifx [1]{%
 \ifx #1\expandafter \@firstoftwo
 \else \expandafter \@secondoftwo
 \fi
}%
\providecommand \natexlab [1]{#1}%
\providecommand \enquote  [1]{``#1''}%
\providecommand \bibnamefont  [1]{#1}%
\providecommand \bibfnamefont [1]{#1}%
\providecommand \citenamefont [1]{#1}%
\providecommand \href@noop [0]{\@secondoftwo}%
\providecommand \href [0]{\begingroup \@sanitize@url \@href}%
\providecommand \@href[1]{\@@startlink{#1}\@@href}%
\providecommand \@@href[1]{\endgroup#1\@@endlink}%
\providecommand \@sanitize@url [0]{\catcode `\\12\catcode `\$12\catcode
  `\&12\catcode `\#12\catcode `\^12\catcode `\_12\catcode `\%12\relax}%
\providecommand \@@startlink[1]{}%
\providecommand \@@endlink[0]{}%
\providecommand \url  [0]{\begingroup\@sanitize@url \@url }%
\providecommand \@url [1]{\endgroup\@href {#1}{\urlprefix }}%
\providecommand \urlprefix  [0]{URL }%
\providecommand \Eprint [0]{\href }%
\providecommand \doibase [0]{http://dx.doi.org/}%
\providecommand \selectlanguage [0]{\@gobble}%
\providecommand \bibinfo  [0]{\@secondoftwo}%
\providecommand \bibfield  [0]{\@secondoftwo}%
\providecommand \translation [1]{[#1]}%
\providecommand \BibitemOpen [0]{}%
\providecommand \bibitemStop [0]{}%
\providecommand \bibitemNoStop [0]{.\EOS\space}%
\providecommand \EOS [0]{\spacefactor3000\relax}%
\providecommand \BibitemShut  [1]{\csname bibitem#1\endcsname}%
\let\auto@bib@innerbib\@empty
\bibitem [{\citenamefont {Hewson}(1993)}]{hewson1993}%
  \BibitemOpen
  \bibfield  {author} {\bibinfo {author} {\bibfnamefont {A.~C.}\ \bibnamefont
  {Hewson}},\ }\href {\doibase 10.1017/CBO9780511470752} {\emph {\bibinfo
  {title} {The Kondo Problem to Heavy Fermions}}},\ Cambridge Studies in
  Magnetism\ (\bibinfo  {publisher} {Cambridge University Press},\ \bibinfo
  {year} {1993})\BibitemShut {NoStop}%
\bibitem [{\citenamefont {Kondo}(1964)}]{Kondo1964}%
  \BibitemOpen
  \bibfield  {author} {\bibinfo {author} {\bibfnamefont {J.}~\bibnamefont
  {Kondo}},\ }\href {\doibase 10.1143/PTP.32.37} {\bibfield  {journal}
  {\bibinfo  {journal} {Progress of Theoretical Physics}\ }\textbf {\bibinfo
  {volume} {32}},\ \bibinfo {pages} {37} (\bibinfo {year} {1964})}\BibitemShut
  {NoStop}%
\bibitem [{\citenamefont {Wilson}(1975)}]{Wilson1975}%
  \BibitemOpen
  \bibfield  {author} {\bibinfo {author} {\bibfnamefont {K.~G.}\ \bibnamefont
  {Wilson}},\ }\href {\doibase 10.1103/RevModPhys.47.773} {\bibfield  {journal}
  {\bibinfo  {journal} {Rev. Mod. Phys.}\ }\textbf {\bibinfo {volume} {47}},\
  \bibinfo {pages} {773} (\bibinfo {year} {1975})}\BibitemShut {NoStop}%
\bibitem [{\citenamefont {Madhavan}\ \emph {et~al.}(1998)\citenamefont
  {Madhavan}, \citenamefont {Chen}, \citenamefont {Jamneala}, \citenamefont
  {Crommie},\ and\ \citenamefont {Wingreen}}]{Madhavan1998}%
  \BibitemOpen
  \bibfield  {author} {\bibinfo {author} {\bibfnamefont {V.}~\bibnamefont
  {Madhavan}}, \bibinfo {author} {\bibfnamefont {W.}~\bibnamefont {Chen}},
  \bibinfo {author} {\bibfnamefont {T.}~\bibnamefont {Jamneala}}, \bibinfo
  {author} {\bibfnamefont {M.~F.}\ \bibnamefont {Crommie}}, \ and\ \bibinfo
  {author} {\bibfnamefont {N.~S.}\ \bibnamefont {Wingreen}},\ }\href {\doibase
  10.1126/science.280.5363.567} {\bibfield  {journal} {\bibinfo  {journal}
  {Science}\ }\textbf {\bibinfo {volume} {280}},\ \bibinfo {pages} {567}
  (\bibinfo {year} {1998})}\BibitemShut {NoStop}%
\bibitem [{\citenamefont {Cronenwett}\ \emph {et~al.}(1998)\citenamefont
  {Cronenwett}, \citenamefont {Oosterkamp},\ and\ \citenamefont
  {Kouwenhoven}}]{Cronenwett1999}%
  \BibitemOpen
  \bibfield  {author} {\bibinfo {author} {\bibfnamefont {S.~M.}\ \bibnamefont
  {Cronenwett}}, \bibinfo {author} {\bibfnamefont {T.~H.}\ \bibnamefont
  {Oosterkamp}}, \ and\ \bibinfo {author} {\bibfnamefont {L.~P.}\ \bibnamefont
  {Kouwenhoven}},\ }\href {\doibase 10.1126/science.281.5376.540} {\bibfield
  {journal} {\bibinfo  {journal} {Science}\ }\textbf {\bibinfo {volume}
  {281}},\ \bibinfo {pages} {540} (\bibinfo {year} {1998})}\BibitemShut
  {NoStop}%
\bibitem [{\citenamefont {Park}\ \emph {et~al.}(2002)\citenamefont {Park},
  \citenamefont {Pasupathy}, \citenamefont {Goldsmith}, \citenamefont {Chang},
  \citenamefont {Yaish}, \citenamefont {Petta}, \citenamefont {Rinkoski},
  \citenamefont {Sethna}, \citenamefont {Abruna}, \citenamefont {McEuen},\ and\
  \citenamefont {Ralph}}]{Park2002}%
  \BibitemOpen
  \bibfield  {author} {\bibinfo {author} {\bibfnamefont {J.}~\bibnamefont
  {Park}}, \bibinfo {author} {\bibfnamefont {A.~N.}\ \bibnamefont {Pasupathy}},
  \bibinfo {author} {\bibfnamefont {J.~I.}\ \bibnamefont {Goldsmith}}, \bibinfo
  {author} {\bibfnamefont {C.}~\bibnamefont {Chang}}, \bibinfo {author}
  {\bibfnamefont {Y.}~\bibnamefont {Yaish}}, \bibinfo {author} {\bibfnamefont
  {J.~R.}\ \bibnamefont {Petta}}, \bibinfo {author} {\bibfnamefont
  {M.}~\bibnamefont {Rinkoski}}, \bibinfo {author} {\bibfnamefont {J.~P.}\
  \bibnamefont {Sethna}}, \bibinfo {author} {\bibfnamefont {H.~D.}\
  \bibnamefont {Abruna}}, \bibinfo {author} {\bibfnamefont {P.~L.}\
  \bibnamefont {McEuen}}, \ and\ \bibinfo {author} {\bibfnamefont {D.~C.}\
  \bibnamefont {Ralph}},\ }\href {https://doi.org/10.1038/nature00791}
  {\bibfield  {journal} {\bibinfo  {journal} {Nature}\ }\textbf {\bibinfo
  {volume} {417}} (\bibinfo {year} {2002})}\BibitemShut {NoStop}%
\bibitem [{\citenamefont {Jarillo-Herrero}\ \emph {et~al.}(2005)\citenamefont
  {Jarillo-Herrero}, \citenamefont {Kong}, \citenamefont {van~der Zant},
  \citenamefont {Dekker}, \citenamefont {Kouwenhoven},\ and\ \citenamefont
  {De~Franceschi}}]{Herrero2005}%
  \BibitemOpen
  \bibfield  {author} {\bibinfo {author} {\bibfnamefont {P.}~\bibnamefont
  {Jarillo-Herrero}}, \bibinfo {author} {\bibfnamefont {J.}~\bibnamefont
  {Kong}}, \bibinfo {author} {\bibfnamefont {H.~S.~J.}\ \bibnamefont {van~der
  Zant}}, \bibinfo {author} {\bibfnamefont {C.}~\bibnamefont {Dekker}},
  \bibinfo {author} {\bibfnamefont {L.~P.}\ \bibnamefont {Kouwenhoven}}, \ and\
  \bibinfo {author} {\bibfnamefont {S.}~\bibnamefont {De~Franceschi}},\ }\href
  {https://www.nature.com/articles/nature03422#supplementary-information}
  {\bibfield  {journal} {\bibinfo  {journal} {Nature}\ }\textbf {\bibinfo
  {volume} {434}} (\bibinfo {year} {2005})}\BibitemShut {NoStop}%
\bibitem [{\citenamefont {Wahl}\ \emph {et~al.}(2005)\citenamefont {Wahl},
  \citenamefont {Diekh\"oner}, \citenamefont {Wittich}, \citenamefont {Vitali},
  \citenamefont {Schneider},\ and\ \citenamefont {Kern}}]{Wahl2005}%
  \BibitemOpen
  \bibfield  {author} {\bibinfo {author} {\bibfnamefont {P.}~\bibnamefont
  {Wahl}}, \bibinfo {author} {\bibfnamefont {L.}~\bibnamefont {Diekh\"oner}},
  \bibinfo {author} {\bibfnamefont {G.}~\bibnamefont {Wittich}}, \bibinfo
  {author} {\bibfnamefont {L.}~\bibnamefont {Vitali}}, \bibinfo {author}
  {\bibfnamefont {M.~A.}\ \bibnamefont {Schneider}}, \ and\ \bibinfo {author}
  {\bibfnamefont {K.}~\bibnamefont {Kern}},\ }\href {\doibase
  10.1103/PhysRevLett.95.166601} {\bibfield  {journal} {\bibinfo  {journal}
  {Phys. Rev. Lett.}\ }\textbf {\bibinfo {volume} {95}},\ \bibinfo {pages}
  {166601} (\bibinfo {year} {2005})}\BibitemShut {NoStop}%
\bibitem [{\citenamefont {N\'eel}\ \emph {et~al.}(2011)\citenamefont {N\'eel},
  \citenamefont {Berndt}, \citenamefont {Kr\"oger}, \citenamefont {Wehling},
  \citenamefont {Lichtenstein},\ and\ \citenamefont {Katsnelson}}]{Neel2011}%
  \BibitemOpen
  \bibfield  {author} {\bibinfo {author} {\bibfnamefont {N.}~\bibnamefont
  {N\'eel}}, \bibinfo {author} {\bibfnamefont {R.}~\bibnamefont {Berndt}},
  \bibinfo {author} {\bibfnamefont {J.}~\bibnamefont {Kr\"oger}}, \bibinfo
  {author} {\bibfnamefont {T.~O.}\ \bibnamefont {Wehling}}, \bibinfo {author}
  {\bibfnamefont {A.~I.}\ \bibnamefont {Lichtenstein}}, \ and\ \bibinfo
  {author} {\bibfnamefont {M.~I.}\ \bibnamefont {Katsnelson}},\ }\href
  {\doibase 10.1103/PhysRevLett.107.106804} {\bibfield  {journal} {\bibinfo
  {journal} {Phys. Rev. Lett.}\ }\textbf {\bibinfo {volume} {107}},\ \bibinfo
  {pages} {106804} (\bibinfo {year} {2011})}\BibitemShut {NoStop}%
\bibitem [{\citenamefont {Spinelli}\ \emph {et~al.}(2015)\citenamefont
  {Spinelli}, \citenamefont {Gerrits}, \citenamefont {Toskovic}, \citenamefont
  {Bryant}, \citenamefont {Ternes},\ and\ \citenamefont {Otte}}]{Spinelli2015}%
  \BibitemOpen
  \bibfield  {author} {\bibinfo {author} {\bibfnamefont {A.}~\bibnamefont
  {Spinelli}}, \bibinfo {author} {\bibfnamefont {M.}~\bibnamefont {Gerrits}},
  \bibinfo {author} {\bibfnamefont {R.}~\bibnamefont {Toskovic}}, \bibinfo
  {author} {\bibfnamefont {B.}~\bibnamefont {Bryant}}, \bibinfo {author}
  {\bibfnamefont {M.}~\bibnamefont {Ternes}}, \ and\ \bibinfo {author}
  {\bibfnamefont {A.~F.}\ \bibnamefont {Otte}},\ }\href
  {http://dx.doi.org/10.1038/ncomms10046} {\bibfield  {journal} {\bibinfo
  {journal} {Nature Communications}\ }\textbf {\bibinfo {volume} {6}} (\bibinfo
  {year} {2015})}\BibitemShut {NoStop}%
\bibitem [{\citenamefont {Toskovic}\ \emph {et~al.}(2016)\citenamefont
  {Toskovic}, \citenamefont {van~den Berg}, \citenamefont {Spinelli},
  \citenamefont {Eliens}, \citenamefont {van~den Toorn}, \citenamefont
  {Bryant}, \citenamefont {Caux},\ and\ \citenamefont {Otte}}]{Toskovic2016}%
  \BibitemOpen
  \bibfield  {author} {\bibinfo {author} {\bibfnamefont {R.}~\bibnamefont
  {Toskovic}}, \bibinfo {author} {\bibfnamefont {R.}~\bibnamefont {van~den
  Berg}}, \bibinfo {author} {\bibfnamefont {A.}~\bibnamefont {Spinelli}},
  \bibinfo {author} {\bibfnamefont {I.~S.}\ \bibnamefont {Eliens}}, \bibinfo
  {author} {\bibfnamefont {B.}~\bibnamefont {van~den Toorn}}, \bibinfo {author}
  {\bibfnamefont {B.}~\bibnamefont {Bryant}}, \bibinfo {author} {\bibfnamefont
  {J.~S.}\ \bibnamefont {Caux}}, \ and\ \bibinfo {author} {\bibfnamefont
  {A.~F.}\ \bibnamefont {Otte}},\ }\href {http://dx.doi.org/10.1038/nphys3722}
  {\bibfield  {journal} {\bibinfo  {journal} {Nature Physics}\ }\textbf
  {\bibinfo {volume} {12}} (\bibinfo {year} {2016})}\BibitemShut {NoStop}%
\bibitem [{\citenamefont {Otte}\ \emph {et~al.}(2008)\citenamefont {Otte},
  \citenamefont {Ternes}, \citenamefont {von Bergmann}, \citenamefont {Loth},
  \citenamefont {Brune}, \citenamefont {Lutz}, \citenamefont {Hirjibehedin},\
  and\ \citenamefont {Heinrich}}]{Otte2008}%
  \BibitemOpen
  \bibfield  {author} {\bibinfo {author} {\bibfnamefont {A.~F.}\ \bibnamefont
  {Otte}}, \bibinfo {author} {\bibfnamefont {M.}~\bibnamefont {Ternes}},
  \bibinfo {author} {\bibfnamefont {K.}~\bibnamefont {von Bergmann}}, \bibinfo
  {author} {\bibfnamefont {S.}~\bibnamefont {Loth}}, \bibinfo {author}
  {\bibfnamefont {H.}~\bibnamefont {Brune}}, \bibinfo {author} {\bibfnamefont
  {C.~P.}\ \bibnamefont {Lutz}}, \bibinfo {author} {\bibfnamefont {C.~F.}\
  \bibnamefont {Hirjibehedin}}, \ and\ \bibinfo {author} {\bibfnamefont
  {A.~J.}\ \bibnamefont {Heinrich}},\ }\href
  {http://dx.doi.org/10.1038/nphys1072} {\bibfield  {journal} {\bibinfo
  {journal} {Nature Physics}\ }\textbf {\bibinfo {volume} {4}},\ \bibinfo
  {pages} {847 EP } (\bibinfo {year} {2008})}\BibitemShut {NoStop}%
\bibitem [{\citenamefont {Blankenbecler}\ \emph {et~al.}(1981)\citenamefont
  {Blankenbecler}, \citenamefont {Scalapino},\ and\ \citenamefont
  {Sugar}}]{Blankenbecler81}%
  \BibitemOpen
  \bibfield  {author} {\bibinfo {author} {\bibfnamefont {R.}~\bibnamefont
  {Blankenbecler}}, \bibinfo {author} {\bibfnamefont {D.~J.}\ \bibnamefont
  {Scalapino}}, \ and\ \bibinfo {author} {\bibfnamefont {R.~L.}\ \bibnamefont
  {Sugar}},\ }\href {\doibase 10.1103/PhysRevD.24.2278} {\bibfield  {journal}
  {\bibinfo  {journal} {Phys. Rev. D}\ }\textbf {\bibinfo {volume} {24}},\
  \bibinfo {pages} {2278} (\bibinfo {year} {1981})}\BibitemShut {NoStop}%
\bibitem [{\citenamefont {White}\ \emph {et~al.}(1989)\citenamefont {White},
  \citenamefont {Scalapino}, \citenamefont {Sugar}, \citenamefont {Loh},
  \citenamefont {Gubernatis},\ and\ \citenamefont {Scalettar}}]{White89}%
  \BibitemOpen
  \bibfield  {author} {\bibinfo {author} {\bibfnamefont {S.~R.}\ \bibnamefont
  {White}}, \bibinfo {author} {\bibfnamefont {D.~J.}\ \bibnamefont
  {Scalapino}}, \bibinfo {author} {\bibfnamefont {R.~L.}\ \bibnamefont
  {Sugar}}, \bibinfo {author} {\bibfnamefont {E.~Y.}\ \bibnamefont {Loh}},
  \bibinfo {author} {\bibfnamefont {J.~E.}\ \bibnamefont {Gubernatis}}, \ and\
  \bibinfo {author} {\bibfnamefont {R.~T.}\ \bibnamefont {Scalettar}},\ }\href
  {\doibase 10.1103/PhysRevB.40.506} {\bibfield  {journal} {\bibinfo  {journal}
  {Phys. Rev. B}\ }\textbf {\bibinfo {volume} {40}},\ \bibinfo {pages} {506}
  (\bibinfo {year} {1989})}\BibitemShut {NoStop}%
\bibitem [{\citenamefont {Assaad}\ and\ \citenamefont
  {Evertz}(2008)}]{Assaad08_rev}%
  \BibitemOpen
  \bibfield  {author} {\bibinfo {author} {\bibfnamefont {F.}~\bibnamefont
  {Assaad}}\ and\ \bibinfo {author} {\bibfnamefont {H.}~\bibnamefont
  {Evertz}},\ }in\ \href {\doibase 10.1007/978-3-540-74686-7_10} {\emph
  {\bibinfo {booktitle} {Computational Many-Particle Physics}}},\ \bibinfo
  {series} {Lecture Notes in Physics}, Vol.\ \bibinfo {volume} {739},\ \bibinfo
  {editor} {edited by\ \bibinfo {editor} {\bibfnamefont {H.}~\bibnamefont
  {Fehske}}, \bibinfo {editor} {\bibfnamefont {R.}~\bibnamefont {Schneider}}, \
  and\ \bibinfo {editor} {\bibfnamefont {A.}~\bibnamefont {Wei{\ss}e}}}\
  (\bibinfo  {publisher} {Springer},\ \bibinfo {address} {Berlin Heidelberg},\
  \bibinfo {year} {2008})\ pp.\ \bibinfo {pages} {277--356}\BibitemShut
  {NoStop}%
\bibitem [{\citenamefont {Bercx}\ \emph {et~al.}(2017)\citenamefont {Bercx},
  \citenamefont {Goth}, \citenamefont {Hofmann},\ and\ \citenamefont
  {Assaad}}]{ALF_v1}%
  \BibitemOpen
  \bibfield  {author} {\bibinfo {author} {\bibfnamefont {M.}~\bibnamefont
  {Bercx}}, \bibinfo {author} {\bibfnamefont {F.}~\bibnamefont {Goth}},
  \bibinfo {author} {\bibfnamefont {J.~S.}\ \bibnamefont {Hofmann}}, \ and\
  \bibinfo {author} {\bibfnamefont {F.~F.}\ \bibnamefont {Assaad}},\ }\href
  {\doibase 10.21468/SciPostPhys.3.2.013} {\bibfield  {journal} {\bibinfo
  {journal} {SciPost Phys.}\ }\textbf {\bibinfo {volume} {3}},\ \bibinfo
  {pages} {013} (\bibinfo {year} {2017})}\BibitemShut {NoStop}%
\bibitem [{\citenamefont {Assaad}(1999)}]{Assaad99a}%
  \BibitemOpen
  \bibfield  {author} {\bibinfo {author} {\bibfnamefont {F.~F.}\ \bibnamefont
  {Assaad}},\ }\href {\doibase 10.1103/PhysRevLett.83.796} {\bibfield
  {journal} {\bibinfo  {journal} {Phys. Rev. Lett.}\ }\textbf {\bibinfo
  {volume} {83}},\ \bibinfo {pages} {796} (\bibinfo {year} {1999})}\BibitemShut
  {NoStop}%
\bibitem [{\citenamefont {Sato}\ \emph {et~al.}(2018)\citenamefont {Sato},
  \citenamefont {Assaad},\ and\ \citenamefont {Grover}}]{SatoT17_1}%
  \BibitemOpen
  \bibfield  {author} {\bibinfo {author} {\bibfnamefont {T.}~\bibnamefont
  {Sato}}, \bibinfo {author} {\bibfnamefont {F.~F.}\ \bibnamefont {Assaad}}, \
  and\ \bibinfo {author} {\bibfnamefont {T.}~\bibnamefont {Grover}},\ }\href
  {\doibase 10.1103/PhysRevLett.120.107201} {\bibfield  {journal} {\bibinfo
  {journal} {Phys. Rev. Lett.}\ }\textbf {\bibinfo {volume} {120}},\ \bibinfo
  {pages} {107201} (\bibinfo {year} {2018})}\BibitemShut {NoStop}%
\bibitem [{\citenamefont {Assaad}(2002)}]{Assaad2002}%
  \BibitemOpen
  \bibfield  {author} {\bibinfo {author} {\bibfnamefont {F.~F.}\ \bibnamefont
  {Assaad}},\ }\href {\doibase 10.1103/PhysRevB.65.115104} {\bibfield
  {journal} {\bibinfo  {journal} {Phys. Rev. B}\ }\textbf {\bibinfo {volume}
  {65}},\ \bibinfo {pages} {115104} (\bibinfo {year} {2002})}\BibitemShut
  {NoStop}%
\bibitem [{\citenamefont {Figgins}\ and\ \citenamefont
  {Morr}(2010)}]{Figgins2010}%
  \BibitemOpen
  \bibfield  {author} {\bibinfo {author} {\bibfnamefont {J.}~\bibnamefont
  {Figgins}}\ and\ \bibinfo {author} {\bibfnamefont {D.~K.}\ \bibnamefont
  {Morr}},\ }\href {\doibase 10.1103/PhysRevLett.104.187202} {\bibfield
  {journal} {\bibinfo  {journal} {Phys. Rev. Lett.}\ }\textbf {\bibinfo
  {volume} {104}},\ \bibinfo {pages} {187202} (\bibinfo {year}
  {2010})}\BibitemShut {NoStop}%
\bibitem [{\citenamefont {Ternes}(2015)}]{Ternes2015}%
  \BibitemOpen
  \bibfield  {author} {\bibinfo {author} {\bibfnamefont {M.}~\bibnamefont
  {Ternes}},\ }\href {http://stacks.iop.org/1367-2630/17/i=6/a=063016}
  {\bibfield  {journal} {\bibinfo  {journal} {New Journal of Physics}\ }\textbf
  {\bibinfo {volume} {17}},\ \bibinfo {pages} {063016} (\bibinfo {year}
  {2015})}\BibitemShut {NoStop}%
\bibitem [{\citenamefont {Morr}(2017)}]{Morr2017}%
  \BibitemOpen
  \bibfield  {author} {\bibinfo {author} {\bibfnamefont {D.~K.}\ \bibnamefont
  {Morr}},\ }\href {http://stacks.iop.org/0034-4885/80/i=1/a=014502} {\bibfield
   {journal} {\bibinfo  {journal} {Reports on Progress in Physics}\ }\textbf
  {\bibinfo {volume} {80}},\ \bibinfo {pages} {014502} (\bibinfo {year}
  {2017})}\BibitemShut {NoStop}%
\bibitem [{\citenamefont {Raczkowski}\ and\ \citenamefont
  {Assaad}(2019)}]{Raczkowski18}%
  \BibitemOpen
  \bibfield  {author} {\bibinfo {author} {\bibfnamefont {M.}~\bibnamefont
  {Raczkowski}}\ and\ \bibinfo {author} {\bibfnamefont {F.~F.}\ \bibnamefont
  {Assaad}},\ }\href {\doibase 10.1103/PhysRevLett.122.097203} {\bibfield
  {journal} {\bibinfo  {journal} {Phys. Rev. Lett.}\ }\textbf {\bibinfo
  {volume} {122}},\ \bibinfo {pages} {097203} (\bibinfo {year}
  {2019})}\BibitemShut {NoStop}%
\bibitem [{\citenamefont {Costi}(2000)}]{TCosti2000}%
  \BibitemOpen
  \bibfield  {author} {\bibinfo {author} {\bibfnamefont {T.~A.}\ \bibnamefont
  {Costi}},\ }\href {\doibase 10.1103/PhysRevLett.85.1504} {\bibfield
  {journal} {\bibinfo  {journal} {Phys. Rev. Lett.}\ }\textbf {\bibinfo
  {volume} {85}},\ \bibinfo {pages} {1504} (\bibinfo {year}
  {2000})}\BibitemShut {NoStop}%
\bibitem [{\citenamefont {Beach}(2004)}]{KBeach2004}%
  \BibitemOpen
  \bibfield  {author} {\bibinfo {author} {\bibfnamefont {K.~S.~D.}\
  \bibnamefont {Beach}},\ }\href {https://arxiv.org/abs/cond-mat/0403055}
  {\bibfield  {journal} {\bibinfo  {journal} {arxiv:0403055}\ } (\bibinfo
  {year} {2004})}\BibitemShut {NoStop}%
\bibitem [{sup()}]{suppl}%
  \BibitemOpen
  \href@noop {} {\bibinfo  {journal} {For details on the level crossings, the
  effective model, the Kondo scales, the mean-field Hamiltonian of an infinite
  chain, and $dI/dV$ spectrum of STM experiment, see Supplemental Material}\
  }\BibitemShut {NoStop}%
\bibitem [{\citenamefont {Luitz}\ \emph {et~al.}(2012)\citenamefont {Luitz},
  \citenamefont {Assaad}, \citenamefont {Novotn\'y}, \citenamefont {Karrasch},\
  and\ \citenamefont {Meden}}]{Luitz11}%
  \BibitemOpen
\bibfield  {journal} {  }\bibfield  {author} {\bibinfo {author} {\bibfnamefont
  {D.~J.}\ \bibnamefont {Luitz}}, \bibinfo {author} {\bibfnamefont {F.~F.}\
  \bibnamefont {Assaad}}, \bibinfo {author} {\bibfnamefont {T.}~\bibnamefont
  {Novotn\'y}}, \bibinfo {author} {\bibfnamefont {C.}~\bibnamefont {Karrasch}},
  \ and\ \bibinfo {author} {\bibfnamefont {V.}~\bibnamefont {Meden}},\ }\href
  {\doibase 10.1103/PhysRevLett.108.227001} {\bibfield  {journal} {\bibinfo
  {journal} {Phys. Rev. Lett.}\ }\textbf {\bibinfo {volume} {108}},\ \bibinfo
  {pages} {227001} (\bibinfo {year} {2012})}\BibitemShut {NoStop}%
\bibitem [{\citenamefont {Karrasch}\ \emph {et~al.}(2010)\citenamefont
  {Karrasch}, \citenamefont {Meden},\ and\ \citenamefont
  {Sch\"onhammer}}]{Karrasch10}%
  \BibitemOpen
  \bibfield  {author} {\bibinfo {author} {\bibfnamefont {C.}~\bibnamefont
  {Karrasch}}, \bibinfo {author} {\bibfnamefont {V.}~\bibnamefont {Meden}}, \
  and\ \bibinfo {author} {\bibfnamefont {K.}~\bibnamefont {Sch\"onhammer}},\
  }\href {\doibase 10.1103/PhysRevB.82.125114} {\bibfield  {journal} {\bibinfo
  {journal} {Phys. Rev. B}\ }\textbf {\bibinfo {volume} {82}},\ \bibinfo
  {pages} {125114} (\bibinfo {year} {2010})}\BibitemShut {NoStop}%
\bibitem [{\citenamefont {Monien}(2010)}]{Monien10}%
  \BibitemOpen
  \bibfield  {author} {\bibinfo {author} {\bibfnamefont {H.}~\bibnamefont
  {Monien}},\ }\href@noop {} {\bibfield  {journal} {\bibinfo  {journal} {Math.
  Comp.}\ }\textbf {\bibinfo {volume} {79}},\ \bibinfo {pages} {857} (\bibinfo
  {year} {2010})}\BibitemShut {NoStop}%
\bibitem [{\citenamefont {Hirsch}\ and\ \citenamefont
  {Fye}(1986)}]{Hirsch1986}%
  \BibitemOpen
  \bibfield  {author} {\bibinfo {author} {\bibfnamefont {J.~E.}\ \bibnamefont
  {Hirsch}}\ and\ \bibinfo {author} {\bibfnamefont {R.~M.}\ \bibnamefont
  {Fye}},\ }\href {\doibase 10.1103/PhysRevLett.56.2521} {\bibfield  {journal}
  {\bibinfo  {journal} {Phys. Rev. Lett.}\ }\textbf {\bibinfo {volume} {56}},\
  \bibinfo {pages} {2521} (\bibinfo {year} {1986})}\BibitemShut {NoStop}%
\bibitem [{\citenamefont {von Bergmann}\ \emph {et~al.}(2015)\citenamefont {von
  Bergmann}, \citenamefont {Ternes}, \citenamefont {Loth}, \citenamefont
  {Lutz},\ and\ \citenamefont {Heinrich}}]{von2015}%
  \BibitemOpen
  \bibfield  {author} {\bibinfo {author} {\bibfnamefont {K.}~\bibnamefont {von
  Bergmann}}, \bibinfo {author} {\bibfnamefont {M.}~\bibnamefont {Ternes}},
  \bibinfo {author} {\bibfnamefont {S.}~\bibnamefont {Loth}}, \bibinfo {author}
  {\bibfnamefont {C.~P.}\ \bibnamefont {Lutz}}, \ and\ \bibinfo {author}
  {\bibfnamefont {A.~J.}\ \bibnamefont {Heinrich}},\ }\href {\doibase
  10.1103/PhysRevLett.114.076601} {\bibfield  {journal} {\bibinfo  {journal}
  {Phys. Rev. Lett.}\ }\textbf {\bibinfo {volume} {114}},\ \bibinfo {pages}
  {076601} (\bibinfo {year} {2015})}\BibitemShut {NoStop}%
\bibitem [{\citenamefont {Anderson}\ \emph {et~al.}(1970)\citenamefont
  {Anderson}, \citenamefont {Yuval},\ and\ \citenamefont
  {Hamann}}]{Anderson11970}%
  \BibitemOpen
  \bibfield  {author} {\bibinfo {author} {\bibfnamefont {P.~W.}\ \bibnamefont
  {Anderson}}, \bibinfo {author} {\bibfnamefont {G.}~\bibnamefont {Yuval}}, \
  and\ \bibinfo {author} {\bibfnamefont {D.~R.}\ \bibnamefont {Hamann}},\
  }\href {\doibase 10.1103/PhysRevB.1.4464} {\bibfield  {journal} {\bibinfo
  {journal} {Phys. Rev. B}\ }\textbf {\bibinfo {volume} {1}},\ \bibinfo {pages}
  {4464} (\bibinfo {year} {1970})}\BibitemShut {NoStop}%
\bibitem [{\citenamefont {Anderson}(1970)}]{Anderson1970}%
  \BibitemOpen
  \bibfield  {author} {\bibinfo {author} {\bibfnamefont {P.~W.}\ \bibnamefont
  {Anderson}},\ }\href {http://stacks.iop.org/0022-3719/3/i=12/a=008}
  {\bibfield  {journal} {\bibinfo  {journal} {Journal of Physics C: Solid State
  Physics}\ }\textbf {\bibinfo {volume} {3}},\ \bibinfo {pages} {2436}
  (\bibinfo {year} {1970})}\BibitemShut {NoStop}%
\bibitem [{\citenamefont {Yosida}(1991)}]{Yosida1991}%
  \BibitemOpen
  \bibfield  {author} {\bibinfo {author} {\bibfnamefont {K.}~\bibnamefont
  {Yosida}},\ }\href@noop {} {\emph {\bibinfo {title} {Theory of Magnetism}}},\
  Springer series in solid sciences\ (\bibinfo  {publisher} {Springer-Verlag},\
  \bibinfo {year} {1991})\BibitemShut {NoStop}%
\bibitem [{\citenamefont {Romeike}\ \emph {et~al.}(2006)\citenamefont
  {Romeike}, \citenamefont {Wegewijs}, \citenamefont {Hofstetter},\ and\
  \citenamefont {Schoeller}}]{Romeike2006}%
  \BibitemOpen
  \bibfield  {author} {\bibinfo {author} {\bibfnamefont {C.}~\bibnamefont
  {Romeike}}, \bibinfo {author} {\bibfnamefont {M.~R.}\ \bibnamefont
  {Wegewijs}}, \bibinfo {author} {\bibfnamefont {W.}~\bibnamefont
  {Hofstetter}}, \ and\ \bibinfo {author} {\bibfnamefont {H.}~\bibnamefont
  {Schoeller}},\ }\href {\doibase 10.1103/PhysRevLett.96.196601} {\bibfield
  {journal} {\bibinfo  {journal} {Phys. Rev. Lett.}\ }\textbf {\bibinfo
  {volume} {96}},\ \bibinfo {pages} {196601} (\bibinfo {year}
  {2006})}\BibitemShut {NoStop}%
\bibitem [{\citenamefont {\ifmmode~\check{Z}\else \v{Z}\fi{}itko}\ \emph
  {et~al.}(2008)\citenamefont {\ifmmode~\check{Z}\else \v{Z}\fi{}itko},
  \citenamefont {Peters},\ and\ \citenamefont {Pruschke}}]{Rok2008}%
  \BibitemOpen
  \bibfield  {author} {\bibinfo {author} {\bibfnamefont {R.}~\bibnamefont
  {\ifmmode~\check{Z}\else \v{Z}\fi{}itko}}, \bibinfo {author} {\bibfnamefont
  {R.}~\bibnamefont {Peters}}, \ and\ \bibinfo {author} {\bibfnamefont
  {T.}~\bibnamefont {Pruschke}},\ }\href {\doibase 10.1103/PhysRevB.78.224404}
  {\bibfield  {journal} {\bibinfo  {journal} {Phys. Rev. B}\ }\textbf {\bibinfo
  {volume} {78}},\ \bibinfo {pages} {224404} (\bibinfo {year}
  {2008})}\BibitemShut {NoStop}%
\end{thebibliography}%
\newpage
\widetext
\begin{center}
\Large{\bf Supplemental Material for: Exploring the Kondo effect of an extended impurity with chains of Co adatoms in a magnetic field}
\end{center}
\subsection{Effective models at level crossings}\label{H_effL1234}
\begin{figure}[htbp]
\centering
\includegraphics[width=\textwidth]{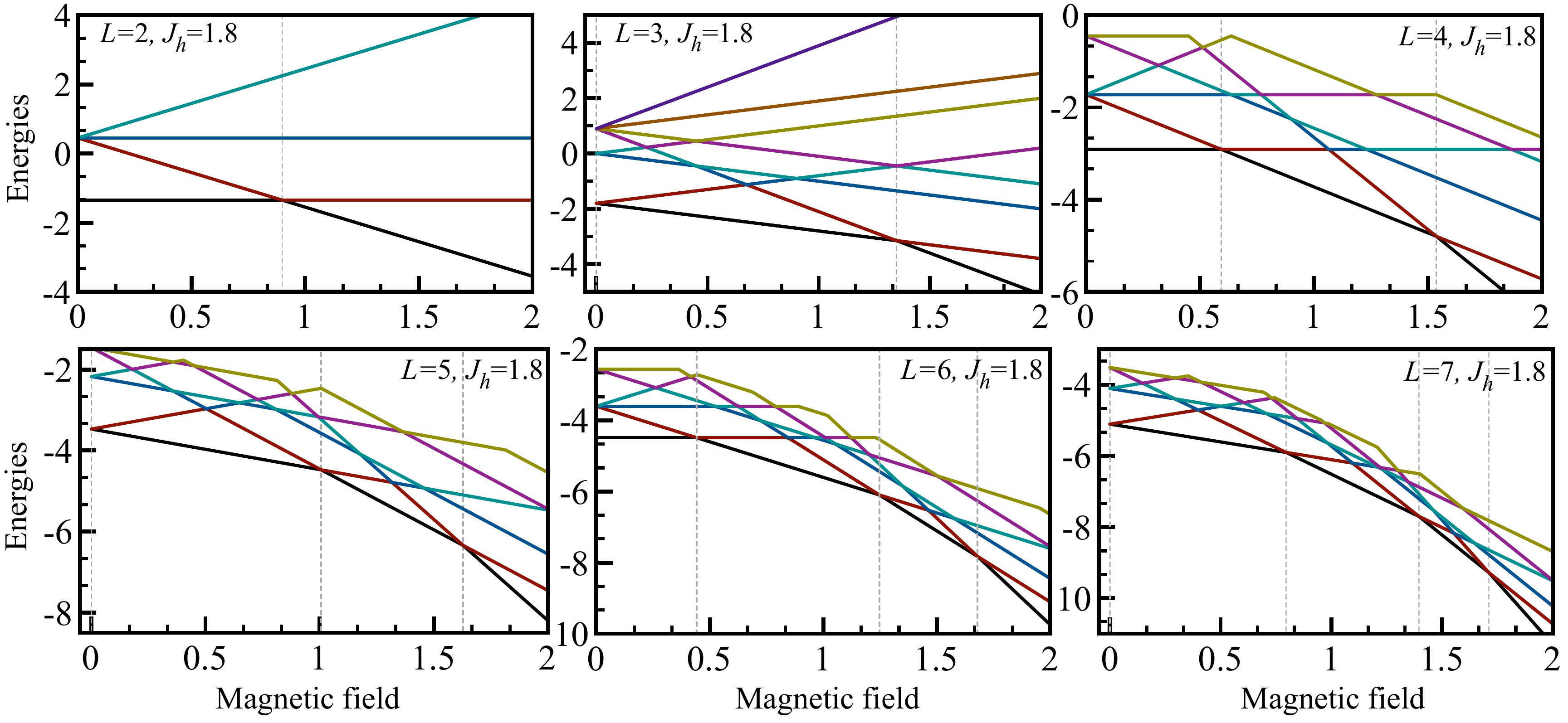}
\caption{Lowest  eigen energies as a function of magnetic field (in $z$-direction in units of $g \mu_B$) for a spin-1/2 Heisenberg chain.}
\label{fig:eigenspectrum_vs_hz}
\includegraphics[width=\textwidth]{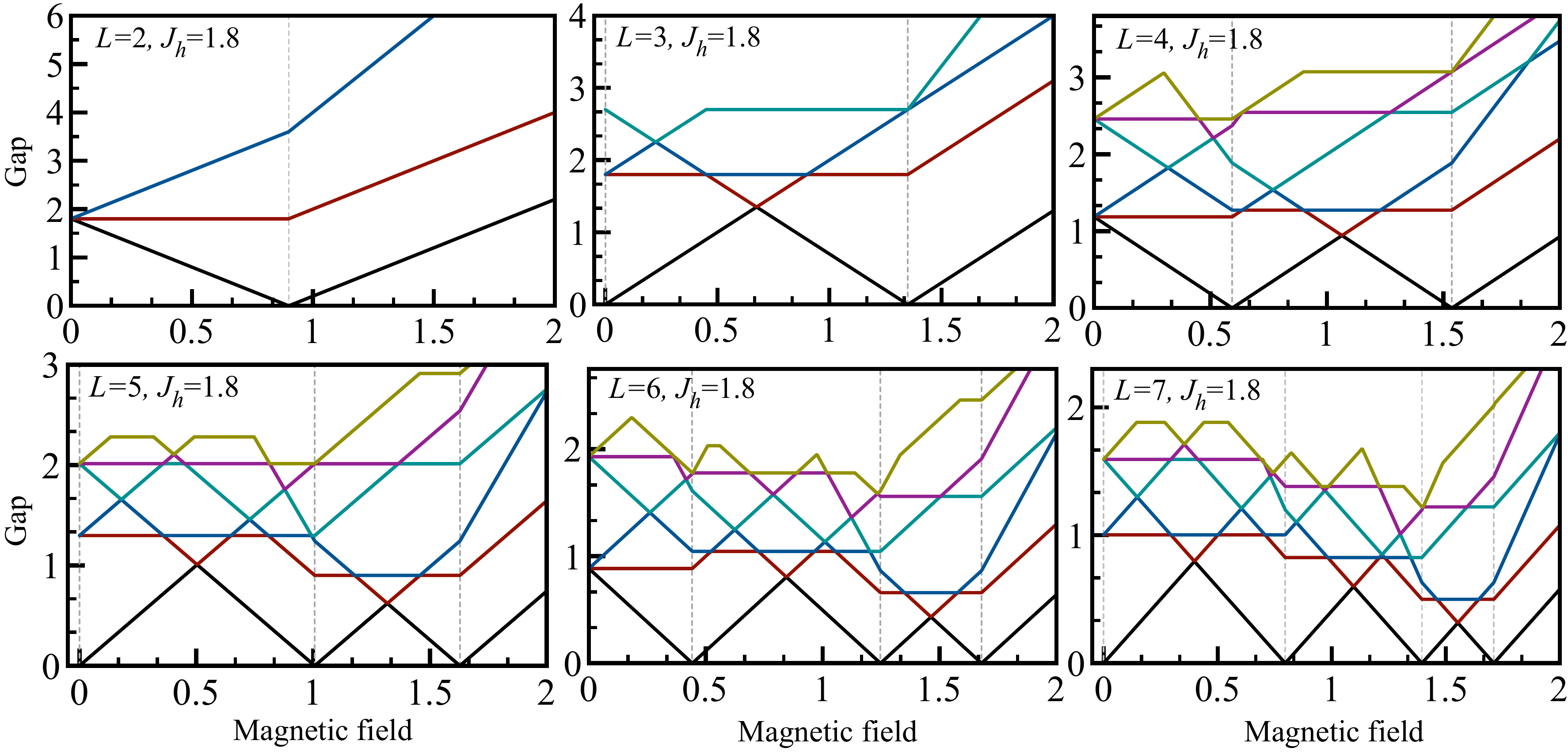}
\caption{Lowest energy excitations as a function of  magnetic field (in $z$-direction  in units of $g \mu_B$) for a spin-1/2 Heisenberg chain.}
\label{fig:gapSU2_vs_hz}
\end{figure}

To derive the effective model at  level crossing $p$ (see Fig.~\ref{fig:eigenspectrum_vs_hz} and Fig.~\ref{fig:gapSU2_vs_hz}) we  project   the Hamiltonian of Eq.~(\ref{model_ham})  onto the  Hilbert space  spanned by the states  $\big\{|\mbf\rangle, |\mmbf\rangle\big\}$  of the  level crossing of a finite Heisenberg chain  at a given critical magnetic field. 
 \begin{eqnarray}
\hat H^{eff}_p=-t\sum_{\langle \i,\j\rangle, \sigma}(\hat c^{\dagger}_{\i,\sigma} \hat c_{\j,\sigma}+h.c)+ J_k\sum^L_{\ve{l}=1} \sum_{\mibf \mjbf=\mbf,\mmbf} \Sbf^c_{\ve{l}} \cdot |\mjbf\rangle \langle \mjbf |\Sbf_{\ve{l}}|\mibf\rangle \langle \mibf|\nonumber\\+(e_{\mbf,p}-\Sbf^z_{\mbf,p} h^z) |\mbf\rangle\langle \mbf|+(e_{\mmbf,p}-\Sbf^z_{\mmbf,p} h^z) |\mmbf\rangle\langle \mmbf|.
 \label{HL1234P}
\end{eqnarray}
The states are   explicitly given  in Table~\ref{tab:exactm1m2}.  
\begin{table}[htbp]
\caption{\label{tab:exactm1m2} The subspace  $\{|\mbf\rangle=|\Sbf_\mbf\Sbf^z_\mbf\rangle_p,|\mmbf \rangle = | \Sbf_\mmbf\Sbf^z_\mmbf\rangle_p\}$ at a  level crossing $p$  of a finite Heisenberg chain ($J_h\sum^{L-1}_{\ve{l}=1} \hat{\ve{S}}_{l} \cdot \hat{\ve{S}}_{\ve{l}+1}$) and corresponding eigenenergies, lowest  excitation gap,  and effective magnetic field.}\begin{ruledtabular}
\begin{tabular}{l | l}
$L=2$& $|0\rangle=\frac{1}{\sqrt{2}}\left(\big|\uparrow\downarrow\rangle-|\downarrow\uparrow\big\rangle\right)$,\quad \quad $\big|1\big\rangle=\big|\uparrow \uparrow\big\rangle$\\
& $e_{0,1}=-\frac{3}{4}J_h$, \quad $e_{1,1}=\frac{J_h}{4}$, \quad $\Delta^{0,1}_1=J_h$, \quad $\mu_{1(2),1}=\frac{1}{4}$
\\\\
$L=3$& $\big|\frac{1}{2}\big\rangle=\frac{1}{\sqrt{6}}\big(|\downarrow\uparrow\uparrow\rangle-2|\uparrow\downarrow\uparrow\rangle+\uparrow\uparrow\downarrow\rangle\big)$, \quad $\big|-\frac{1}{2}\big\rangle=\frac{1}{\sqrt{6}}\left(|\downarrow\downarrow\uparrow\rangle-2|\downarrow\uparrow\downarrow\rangle+\uparrow\downarrow\downarrow\rangle\right)$\\
&  $e_{\frac{1}{2},1}=-J_h$, \quad $e_{-\frac{1}{2},1}=-J_h$, \quad $\Delta^{-\frac{1}{2},\frac{1}{2}}_1=0$, \quad $\mu_{1(3),1}=0$, \quad   $\mu_{2,1}=0$\\\\
$L=3$& $\big|\frac{1}{2}\big\rangle=\frac{1}{\sqrt{6}}\big(\big|\downarrow\uparrow\uparrow\big\rangle-2\big|\uparrow\downarrow\uparrow\big\rangle+\big|\uparrow\uparrow\downarrow\big\rangle\big)$,\quad \quad $\big|\frac{3}{2}\big\rangle=\big|\uparrow \uparrow\uparrow\big\rangle$\\
& $e_{\frac{1}{2},2}=-J_h$, \quad $e_{\frac{3}{2},2}=\frac{J_h}{2}$, \quad $\Delta^{\frac{1}{2},\frac{3}{2}}_2=\frac{3}{2}J_h$, \quad $\mu_{1(3),2}=\frac{5}{12}$, \quad $\mu_{2,2}=\frac{1}{6}$\\\\
$L=4$ & $\big|0\big\rangle=\frac{1}{\sqrt{6}}\Big[\frac{1}{(1+\sqrt{3})}\big(\big|\downarrow\downarrow\uparrow\uparrow\big\rangle+\big|\uparrow\uparrow\downarrow\downarrow\big\rangle\big)+\big|\downarrow\uparrow\uparrow\downarrow\big\rangle+\big|\uparrow\downarrow\downarrow\uparrow\big\rangle-\sqrt{1+\frac{\sqrt{3}}{2}}\big(\big|\downarrow\uparrow\downarrow\uparrow\big\rangle+\big|\uparrow\downarrow\uparrow\downarrow\big\rangle\big)\Big]$\\
&$\big|1\big\rangle=\frac{1}{2\sqrt{2+\sqrt{2}}}\Big[-\big|\downarrow\uparrow\uparrow\uparrow\big\rangle+\big|\uparrow\uparrow\uparrow\downarrow\big\rangle+(1+\sqrt{2})\big(\big|\uparrow\downarrow\uparrow\uparrow\big\rangle-|\uparrow\uparrow\downarrow\uparrow\big\rangle\big)\Big]$\\
&$e_{0,1}=\frac{J_h}{4}\big(-3-2\sqrt{3}\big)$,\quad $e_{1,1}=\frac{J_h}{4}\big(-1-2\sqrt{2}\big)$, \quad $\Delta^{0,1}_1=\frac{J_h}{2} (1-\sqrt{2}+\sqrt{3})$,\quad $\mu_{1(4),1}=\frac{(2+\sqrt{2})}{16}$,\quad $\mu_{2(3),1}=\frac{(2-\sqrt{2})}{16}$\\\\
$L=4$&$\big|1\big\rangle=\frac{1}{2\sqrt{2+\sqrt{2}}}\Big[-\big|\downarrow\uparrow\uparrow\uparrow\big\rangle+\big|\uparrow\uparrow\uparrow\downarrow\big\rangle+(1+\sqrt{2})\big(\big|\uparrow\downarrow\uparrow\uparrow\big\rangle-\big|\uparrow\uparrow\downarrow\uparrow\big\rangle\big)\Big]$, \quad \quad  $\big|2\big\rangle=\big|\uparrow\uparrow\uparrow\uparrow\big\rangle$\\
& $e_{1,2}=\frac{J_h}{4}\big(-1-2\sqrt{2}\big)$,\quad $e_{2,2}=\frac{3J_h}{4}$, \quad $\Delta^{1,2}_2=J_h\big(1+\frac{1}{\sqrt{2}}\big)$, \quad $\mu_{1(4),2}=\frac{(6+\sqrt{2})}{16}$,\quad $\mu_{2(3),2}=\frac{(6-\sqrt{2})}{16}$
\end{tabular}
\end{ruledtabular}
\end{table}
We further introduce   pseudo spin-1/2 operators; $\hat{\tau}^x=|\mmbf\rangle\langle \mbf|+|\mbf\rangle\langle \mmbf|$, $\hat {\tau}^y=-i(|\mmbf\rangle\langle \mbf|-|\mbf\rangle\langle \mmbf|)$ and $\hat{\tau}^z=|\mmbf\rangle \langle \mmbf|-|\mbf\rangle \langle \mbf|$  and impose the constraint $\identity=|\mbf\rangle \langle \mbf|+|\mmbf\rangle \langle \mmbf|$. In terms of the $\ve{\tau}$ operators  Eq.~(\ref{HL1234P})  takes the  form given in  Eq.~(\ref{effe_ham}) with a set of site dependent effective couplings given in Table~\ref{tab:j_x,z}.  The other effective parameters of Eq.~(\ref{effe_ham}) are given in Table~\ref{tab:exactm1m2} up to L=4. When calculating the local matrix elements, we obtain  an alternative $\pm$ sign of $j^{\perp}_{\ve{l},p}$ which does not  affect  Kondo physics~\cite{Anderson11970,Anderson1970}. For simplicity we omitted this sign $(-1)^{\ve{l}}$ in   Table~\ref{tab:j_x,z} of the main text.

\begin{figure}[htbp]
\centering
\includegraphics[width=11.5cm]{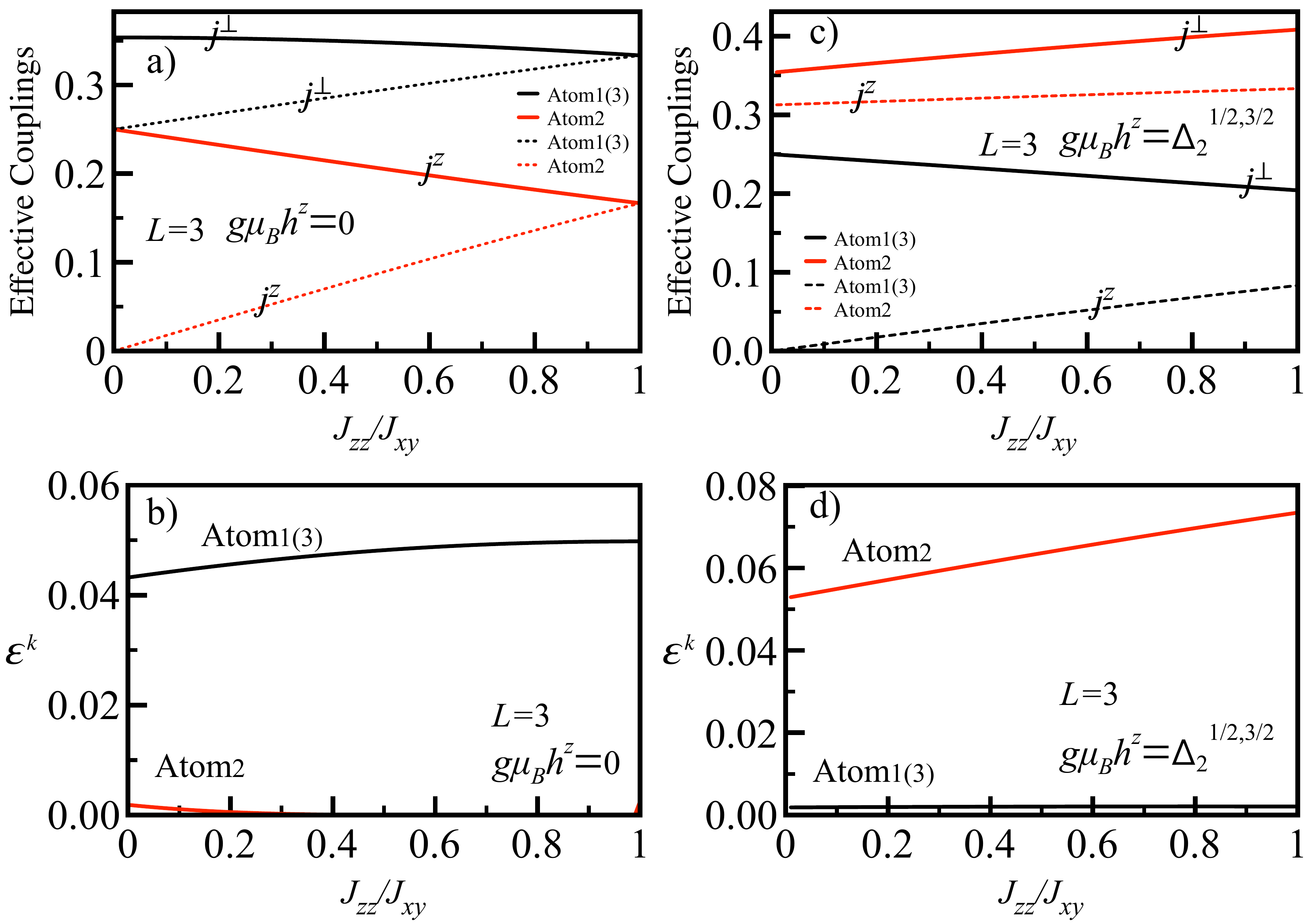}
\caption{Top: Site dependent effective couplings as a function of  $J_{zz}/J_{xy}$ at level crossings for $L=3$. Bottom: Corresponding effective Kondo scale as a function of  $J_{zz}/J_{xy}$. Here, the effective Kondo scale is estimated as; $\epsilon^k_{l(l^\prime),p}\sim e^{-\frac{1} {j_{l,p}}}$  with  $\frac{1}{\ j_{l,p}}= \frac{1}{\sqrt ({j^{\perp}_{l,p})^2-(j^{z}_{l,p})^2 }} \times  \tan^{-1} \Big ( \frac {\sqrt{ (j^{\perp}_{l,p})^2-(j^{z}_{l,p})^2}}{j^{z}_{l,p}}\Big)$ (see Section~\ref{Effec_Kondo_scale}.).}
\label{fig:XY_SU2_Kondoscale_L3}
\includegraphics[width=12.cm]{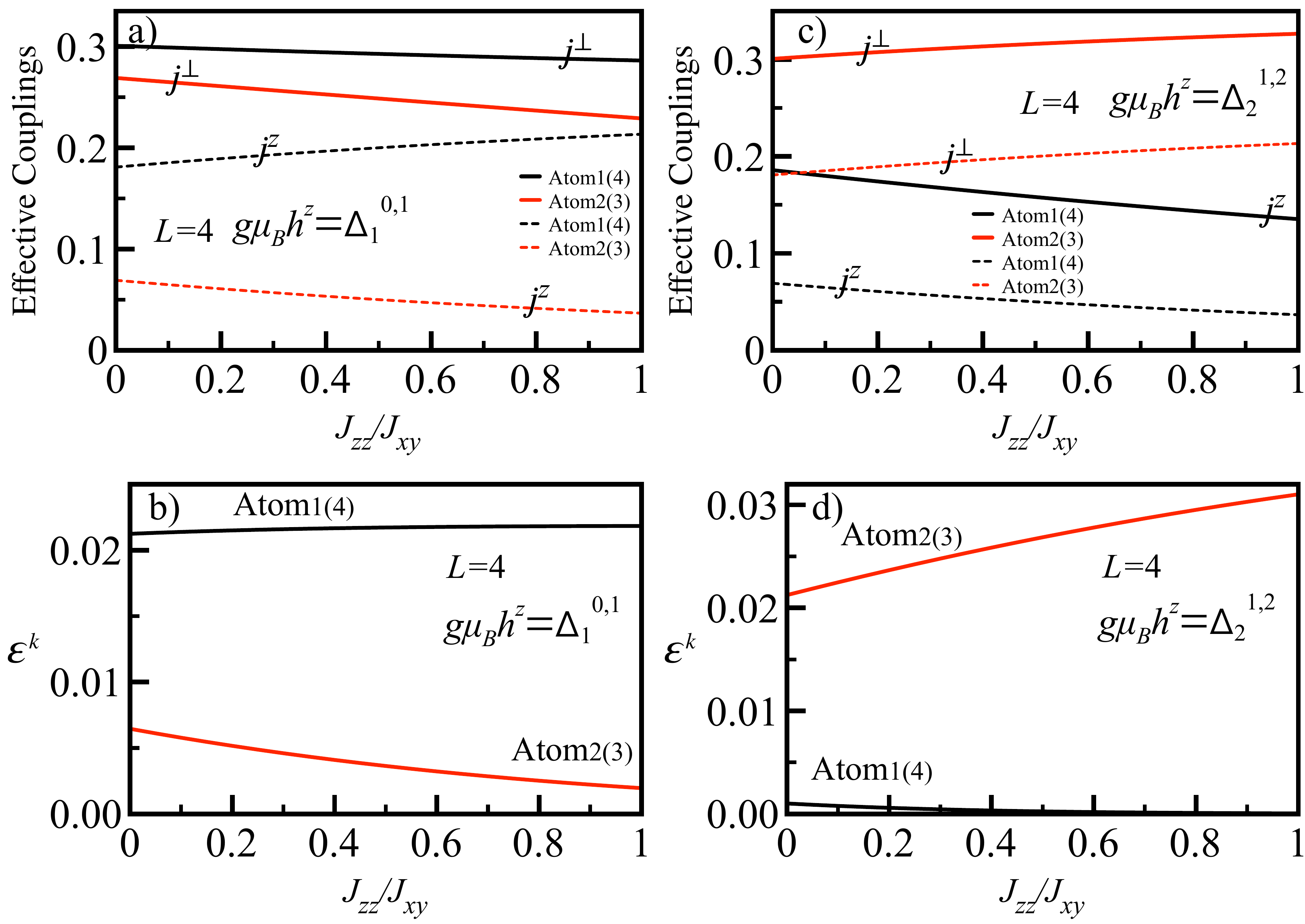}
\caption{Top: Site dependent effective couplings as a function of  $J_{zz}/J_{xy}$ at level crossings for $L=4$. Bottom:  Corresponding effective Kondo scale ($\epsilon^k_{l(l^\prime),p}\sim e^{-\frac{1} {j_{l,p}}}$) as a function of  $J_{zz}/J_{xy}$.}
\label{fig:XY_SU2_Kondoscale_L4}
\end{figure}

Noticeably, for a Heisenberg chain ($J_h\sum^{L-1}_{\ve{l}=1} \hat{\ve{S}}_{\ve{l}} \cdot \hat{\ve{S}}_{\ve{l}+1}$) the effective couplings $(j^{\perp,z}_{\ve{l}(\ve{l}^\prime),p})$ are independent of $J_h$ (see Table~\ref{tab:j_x,z}). However,  for an XXZ chain ($J_{xy}\sum^{L-1}_{\ve{l}=1} (\hat{S^x}_{\ve{l}}\hat{S^x}_{\ve{l}+1}+\hat{S^y}_{\ve{l}}\hat{S^y}_{\ve{l}+1})+J_{zz}\sum^{L-1}_{\ve{l}=1} \hat{S^z}_{\ve{l}}\hat{S^z}_{\ve{l}+1}$) the ratio of the exchange parameters $J_{zz}/J_{xy}$  appears in the expression of the  effective couplings.  Since they do not have a   simple form,  we choose to plot them  as a function of $J_{zz}/J_{xy}$ in Fig.\ref{fig:XY_SU2_Kondoscale_L3} and Fig.\ref{fig:XY_SU2_Kondoscale_L4} for chains of three and four atoms respectively. As for the  Heisenberg chain ($J_{zz}/J_{xy}=1$), there is always a  strong site dependence of the effective couplings when varying $J_{zz}/J_{xy}=0, \cdots,1$.  Hence, the magnetic field induced level crossings in finite XXZ and  Heisenberg chains is expected to show a similar site dependent Kondo physics. 

~

\newpage
\subsection{Effective Kondo Scale at level crossings (Anderson Poor man scaling approach)}  \label{Effec_Kondo_scale}
Starting from the effective Kondo Hamiltonian for a chain of two atoms at  singlet-triplet level crossing, which reads
\begin{eqnarray}
  \Hhat^{eff}_1 & = &e_{0,1}-t\sum_{\langle \i,\j\rangle, \sigma}(\hat c^{\dagger}_{\i,\sigma} \hat c_{\j,\sigma}+h.c)+J_k\Big\{j^{\perp}  \big[(\hat S^{x,c}_{{\ve{l}}_{1}}+\hat S^{x,c}_{{\ve{l}}_{2}})\hat \tau^x+(\hat S^{y,c}_{{\ve{l}}_{1}}+
  \hat S^{y,c}_{{\ve{l}}_{2}})\hat \tau^y\big] \nonumber\\ &&  +j^{z}(\hat S^{z,c}_{{\ve{l}}_{1}}+ \hat S^{z,c}_{{\ve{l}}_{2}}) \hat{\tau}^z +\mu (\hat S^{z,c}_{{\ve{l}}_{1}}+
    \hat S^{z,c}_{{\ve{l}}_{2}}) \Big\} + \frac{1}{2}\big(\Delta^{0,1}_1-g\mu_Bh^z\big)\big (\identity+\hat \tau^z\big)
  \label{HameffL2AS}
\end{eqnarray}
where,  $j^{\perp}=\frac{1}{2\sqrt{2}}$, $j^z=\frac{1}{4}$ and $\mu =\frac{1}{4}$, 
we use the following unitary transformation for conduction electrons in Eq.~(\ref{HameffL2AS}),
\begin{equation}
\begin{aligned}
\hat \S^c_B=\frac{1}{\sqrt{2}}\Big(\hat{\S}^c_{{\ve{l}}_{1}}+\hat{\S}^c_{{\ve{l}}_{2}}\Big), \quad \hat \S^c_A=\frac{1}{\sqrt{2}}\Big(\hat{\S}^c_{{\ve{l}}_{1}}-\hat{\S}^c_{{\ve{l}}_{2}}\Big)
 \label{UT_AS}
 \end{aligned}
\end{equation}
to rewrite it as
 \begin{equation}
\begin{aligned}
  \Hhat^{eff}_1 =e_{0,1}  -t\sum_{\langle \i,\j\rangle, \sigma}(\hat c^{\dagger}_{\i,\sigma} \hat c_{\j,\sigma}+h.c)+ \tilde{j}^{\perp}( \hat {S}^{x,c}_{B}\hat{\tau}^x+\hat{S}^{y,c}_{B}\hat{\tau}^y)+\tilde{j}^{z} \hat{S}^{z,c}_{B} \hat \tau^z +\tilde\mu  \hat{S}^{z,c}_{B}+\frac{1}{2}\big(\Delta^{0,1}_1-g\mu_Bh^z\big)\big (\identity+\hat \tau^z\big)
  \label{HameffAL2}
 \end{aligned}
\end{equation}
where, $\tilde{j}^\perp=\sqrt{2}J_k j^{\perp}$ and $\tilde{j}^z=\sqrt{2}J_k j^{z}$. The  effective  Hamiltonian given in Eq.(\ref{HameffAL2}) corresponds to an  anisotropic  single impurity  Kondo Hamiltonian.   Following Anderson's poor man scaling approach~\cite{Anderson11970,Anderson1970} the effective Kondo scale can be estimated by integrating and solving  two differential equations (see below) obtained from $T$-matrix scattering  process of the conduction electron scattering off the pseudo spin-1/2  degree of freedom,   $\ve{\tau}$:
 \begin{eqnarray}
\frac {d \tilde {j}^{z}}{d \ln D} =2\rho (\tilde {j}^{\perp})^2
\label{dbydjz}
\end{eqnarray}
 and   
 \begin{eqnarray}
 \frac {d \tilde {j}^{\perp}}{d \ln D}=2\rho  \tilde {j}^{\perp} \tilde {j}^{z}
\label{dbydjxy}
 \end{eqnarray}
where $D$ is  the  half bandwidth cutoff and $\rho$ the  density of states at the  Fermi level. In the isotropic case  $\tilde {j}^{\perp}= \tilde {j}^{z}={j}$, the two differential Eqs.  are identical and yield   the Kondo scale; $\epsilon^k \sim e^{-\frac{1}{\rho j}}$. In the anisotropic case $\tilde {j}^{\perp}\ne\tilde {j}^{z}$ the two Eqs.~(\ref{dbydjz}) and.~(\ref{dbydjxy}) give a scaling trajectory $(\tilde{j}^z)^2-(\tilde j^{\perp})^2 =const$, and the effective Kondo scale  can be obtained by the flow of the renormalised couplings along the trajectory. 

\begin{figure}[htbp]
\centering
\includegraphics[width=\textwidth]{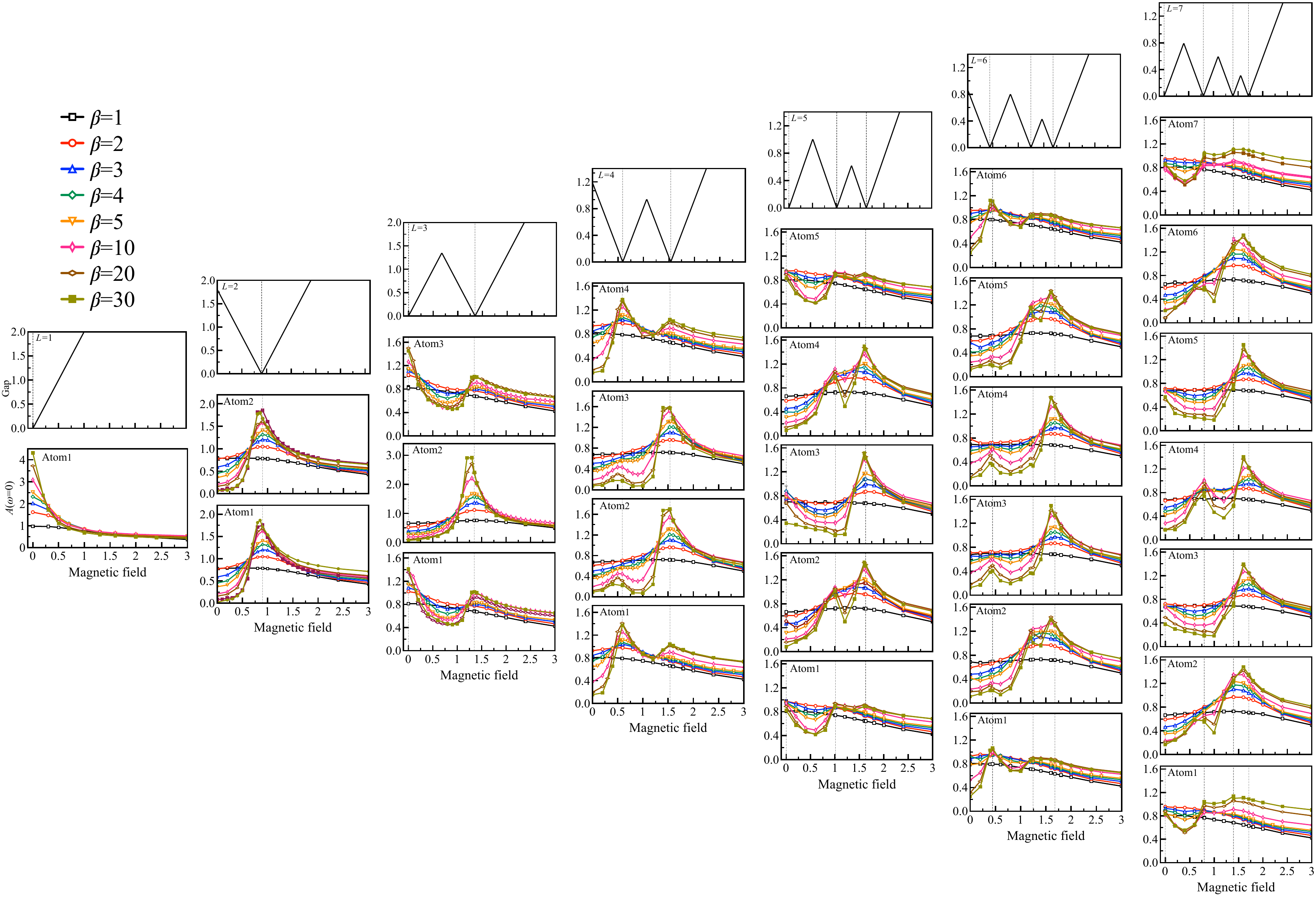}
\caption{QMC results for spectral function at $\omega=0$ as a function of an external magnetic field for $J_k/t=2$, $J_h/t=1.8$ and for different values of inverse temperature $\beta=t/k_BT$ up to $L=7$. This figure is directly comparable with Fig.3 reported in Ref.~[\onlinecite{Toskovic2016}] along a cut on zero bias conductance data around 330 mK.}
\label{fig:spectral_temp}
\end{figure}

For a chain of length $L$   one can write the effective Hamiltonian in terms of $\hat {\S}^c_{B(A)}$  as:
 \begin{equation}
\begin{aligned}
  \Hhat^{eff}_p &=-t\sum_{\langle \i,\j\rangle, \sigma}(\hat c^{\dagger}_{\i,\sigma} \hat c_{\j,\sigma}+h.c)+cJ_k \sum_{\ve{l}} \Big\{ j^{\perp}_{\ve{l},p}( \hat {S}^{x,c}_{\ve{l},B(A)}\hat{\tau}^x+\hat{S}^{y,c}_{\ve{l},B(A)}\hat{\tau}^y)+j^{z}_{\ve{l},p} \hat{S}^{z,c}_{\ve{l},B(A)} \hat \tau^z +\mu_{\ve{l},p}  \hat{S}^{z,c}_{\ve{l},B(A)}\Big\}\\ & +\frac{1}{2}\big(\Delta^{\mbf,\mmbf}_p-g\mu_Bh^z\big)\big (\identity+\hat \tau^z\big).
  \label{HameffASL}
 \end{aligned}
\end{equation}
Here, the summation  $\sum_{\ve{l}}$  goes over $\ve{l}=1, \cdots, \frac{L}{2} (\frac{L+1}{2})$  for  even(odd) $L$, and, $c$ is a normalisation factor arising from the  bonding or antibonding ($B(A)$) selections of conduction electrons involved in the  Kondo effect.    This selection rule stems from  the inversion symmetry present in the Heisenberg chain and involves  pairs of sites  $\ve{l}(=1, \cdots, \frac{L}{2})$ and $\ve{l}^\prime(=L-(\ve{l}-1), \cdots,\frac{L}{2}+1$). One can define a site dependent  effective Kondo scale of Eq.~(\ref{HameffASL})  by considering the flow of renormalised couplings along the scaling trajectories,
\begin{eqnarray}
(\tilde  {j}^z_{\ve{l},p})^2-(\tilde j^{\perp}_{\ve{l},p})^2 =const
\label{scaling_jxjzL}
\end{eqnarray}
where $\tilde{j}^\perp_{\ve{l},p}=cJ_k j^{\perp}_{\ve{l},p}$ and $\tilde{j}^z_{\ve{l},p}=cJ_k j^{z}_{\ve{l},p}$.  To  estimate the  Kondo scale we use  $c=\sqrt{2}$  if two atoms at the position $\ve{l}$ and $\ve{l}^\prime$ are symmetrically involved in the  Kondo resonance and $c=1$ if only one atom  shows a Kondo resonance. The latter case corresponds  for example to the central atom of an odd sized  chain.  Furthermore, we use a constant density of states $\rho=1$ in all cases. 
At a level crossing $p$ and depending on the sign of $\tilde{j}^{z}_{\ve{l},p}$~\cite{Anderson11970,Anderson1970,Yosida1991,Romeike2006,Rok2008} (see Table~\ref{tab:epsilonk_signjz}) a relative estimate of Kondo scale ($\epsilon^k_{\ve{l},p}\sim e^{-\frac{1} { \rho \tilde j_{\ve{l},p}}}$)  is given in Table~\ref{tab:epsilonk_L1234567}. 
\begin{table}[htbp]
\caption{\label{tab:epsilonk_signjz} Kondo scale ($\epsilon^k_{\ve{l},p}\sim e^{-\frac{1} { \rho \tilde j_{\ve{l},p}}}$) according to sign of $\tilde{j}^{z}_{\ve{l},p}$ for an anisotropic Kondo Hamiltonian.} 
\begin{ruledtabular}
\begin{tabular}{l|llll|llll|llll|llll|llll|lll}
&$\tilde{j}^z_{\ve{l},p}$ $>$ 0& \quad\quad\quad\quad\quad$\tilde{j}^z_{\ve{l},p}$ $<$ 0\\
\hline
\hline
$|\tilde{j}^{\perp}_{\ve{l},p}|>|\tilde{j}^z_{\ve{l},p}|$&$\frac{1}{\tilde{j}_{\ve{l},p}}=\frac{1}{\sqrt {({\tilde{j}^{\perp}_{\ve{l},p})^2-(\tilde{j}^{z}_{\ve{l},p})^2 }}} \times  \tan^{-1} \Big (\frac {\sqrt{ (\tilde{j}^{\perp}_{\ve{l},p})^2-(\tilde{j}^{z}_{\ve{l},p})^2}}{\tilde{j}^{z}_{\ve{l},p}} \Big)$& \quad\quad\quad\quad $\frac{1}{\tilde{j}_{\ve{l},p}}=\frac{1}{\sqrt {({\tilde{j}^{\perp}_{\ve{l},p})^2-(\tilde{j}^{z}_{\ve{l},p})^2}}} \times \Big[ \pi+ \tan^{-1} \Big (\frac {\sqrt{ (\tilde{j}^{\perp}_{\ve{l},p})^2-(\tilde{j}^{z}_{\ve{l},p})^2}}{\tilde{j}^{z}_{\ve{l},p}} \Big)\Big]$\\
 \hline
$|\tilde{j}^{\perp}_{\ve{l},p}|<|\tilde{j}^z_{\ve{l},p}|$&$\frac{1}{\tilde{j}_{\ve{l},p}}=\frac{1}{\sqrt {({\tilde{j}^{z}_{\ve{l},p})^2-(\tilde{j}^{\perp}_{\ve{l},p})^2 }}} \times\tanh^{-1} \Big(\frac {\sqrt{(\tilde{j}^{z}_{\ve{l},p})^2-(\tilde{j}^{\perp}_{\ve{l},p})^2}}{\tilde{j}^{z}_{\ve{l},p}} \Big)$ &\quad \quad\quad\quad0\\
\end{tabular}
\end{ruledtabular}
\end{table}
\newpage
\begin{table}[htbp]
\caption{\label{tab:epsilonk_L1234567} Effective Kondo scale $(\epsilon^k_{\ve{l}(\ve{l}^\prime),p}\sim e^{-\frac{1} {\tilde{j}_{\ve{l},p}}})$ at the level crossings for two $J_k=2(1.5)$ values up to L=7 using Anderson poor man scaling approach.  Correspondingly, the Fig.\ref{fig:spectral_temp}  shows  that proper Kondo resonances appears in the QMC simulation at sites where $\epsilon^k_{\ve{l}(\ve{l}^\prime),p}$ dominates.} 
\begin{ruledtabular}
\begin{tabular}{l|llll}
$L=2$ & $\epsilon^{k}_{1(2),1}\sim0.329(0.227)$\\
$L=3$ & $\epsilon^{k}_{1(3),1}\sim 0.346(0.243)$ & $\epsilon^{k}_{2,1}\sim0.049(0.018)$\\
$L=3$ & $\epsilon^{k}_{1(3),2}\sim0.113(0.055)$ & $\epsilon^{k}_{2,2}\sim0.271(0.176)$\\
$L=4$ & $\epsilon^{k}_{1(4),1}\sim0.259(0.165)$ & $\epsilon^{k}_{2(3),1}\sim0.113(0.055)$\\
$L=4$ & $\epsilon^{k}_{1(4),2}\sim0.029(0.009)$ & $\epsilon^{k}_{2(3),2}\sim0.293(0.194)$\\
$L=5$ & $\epsilon^{k}_{1(5),1}\sim0.251(0.159)$ & $\epsilon^{k}_{2(4),1}\sim0.091(0.041)$ & $\epsilon^k_{3,1}\sim0.171(0.095)$\\
$L=5$ & $\epsilon^{k}_{1(5),2}\sim0.124(0.061)$ & $\epsilon^{k}_{2(4),2}\sim0.249(0.157)$ & $\epsilon^{k}_{3,2}\sim0.0004(0.00003)$\\
$L=5$ & $\epsilon^{k}_{1(5),3}\sim0.007(0.001)$ & $\epsilon^{k}_{2(4),3}\sim0.189(0.109)$ & $\epsilon^{k}_{3,3}\sim0.164(0.089)$\\
$L=6$ & $\epsilon^{k}_{1(6),1}\sim0.196(0.114)$ & $\epsilon^{k}_{2(5),1}\sim0.026(0.008)$ & $\epsilon^{k}_{3(4),1}\sim0.151(0.079)$\\
$L=6$ & $\epsilon^{k}_{1(6),2}\sim0.054(0.021)$ & $\epsilon^{k}_{2(5),2}\sim0.251(0.159)$ & $\epsilon^{k}_{3(4),2}\sim0.051(0.019)$\\
$L=6$ & $\epsilon^{k}_{1(6),3}\sim0.002(0.0002)$ & $\epsilon^{k}_{2(5),3}\sim0.113(0.055)$ & $\epsilon^{k}_{3(4),3}\sim0.224(0.136)$\\
$L=7$ & $\epsilon^{k}_{1(7),1}\sim0.186(0.106)$ & $\epsilon^{k}_{2(6),1}\sim0.062(0.025)$ & $\epsilon^{k}_{3(5),1}\sim0.231(0.142)$& $\epsilon^{k}_{4,1}\sim0.035(0.012)$\\
$L=7$ & $\epsilon^{k}_{1(7),2}\sim0.103(0.048)$ & $\epsilon^{k}_{2(6),2}\sim0.119(0.059)$ & $\epsilon^{k}_{3(5),2}\sim0.012(0.003)$& $\epsilon^{k}_{4,2}\sim0.151(0.081)$\\
$L=7$ & $\epsilon^{k}_{1(7),3}\sim0.022(0.006)$ & $\epsilon^{k}_{2(6),3}\sim0.214(0.128)$ & $\epsilon^{k}_{3(5),3}\sim0.123(0.062)$& $\epsilon^{k}_{4,3}\sim0.003(0.0004)$\\
$L=7$ & $\epsilon^{k}_{1(7),4}\sim0.00017(10^{-6})$ & $\epsilon^{k}_{2(6),4}\sim0.063(0.025)$ & $\epsilon^{k}_{3(5),4}\sim0.168(0.093)$& $\epsilon^{k}_{4,4}\sim0.108(0.052)$\\
\end{tabular}
\end{ruledtabular}
\end{table}
A  detailed temperature dependence    of $A_{\ve{l}}(\omega=0)  \simeq \frac{1}{\pi}   \beta G_{\ve{l}}(\tau = \beta/2) $  as a function of magnetic field and chain length  is given  in Fig.~\ref{fig:spectral_temp}.  Upon inspection,  one will see that at a  given critical magnetic field corresponding to a level crossing,  a peak occurs at the site where the local Kondo scale dominates.   
\subsection{Large-N mean field for the infinite Heisenberg chain of adatoms}\label{LargeN_meanfield}
We consider an infinite  Heisenberg chain of adatoms with periodic  boundary conditions. The unit cell, $\ve{l}$,  contains $n \in \left[ 1 \cdots N_c \right] $ conduction electrons  $ \hat{c}_{\ve{l},n,\sigma} $ and a single spin degree of freedom.  In this case,   we can use the   identity  $J_h \ve{S}_{\ve{l}} \cdot \ve{S}_{\ve{l}+1} =  -\frac{J_h}{4}  \left( D^{}_{\ve{l},\ve{l}+1}  D^{\dagger}_{\ve{l},\ve{l}+1}  +  D^{\dagger}_{\ve{l},\ve{l}+1}  D^{}_{\ve{l},\ve{l}+1}  \right) $    with  $\ve{S}_l = \frac{1}{2} \ve{d}^{\dagger}_l  \ve{\sigma} \ve{d}_{\ve{l}}$, $D^{}_{\ve{l},\ve{l}+1}  = \ve{d}^{\dagger}_{\ve{l}}  \ve{d}^{}_{\ve{l}+1} $    and constraint $ \ve{d}^{\dagger}_{\ve{l}}  \ve{d}^{}_{\ve{l}}  =1$  to define the large-N mean-field saddle point of  Eq.~(\ref{model_ham})
\begin{eqnarray} 
 \hat{H}_{MF}  =& & \sum_{n,n',k,\sigma} \hat{c}^{\dagger}_{k,n,\sigma} T(k)_{n,n'} \hat{c}^{}_{k,n,\sigma}  
-\frac{J_h\chi} {4} \sum_{\ve{l},\sigma} (  \hat{d}^{\dagger}_{\ve{l},\sigma} \hat{d}^{}_{{\ve{l}}+1,\sigma}+ h.c.)     - \frac{1}{2}g \mu_B h^z 
\sum_{l,\sigma} \sigma \hat{d}^{\dagger}_{\ve{l},\sigma} \hat{d}^{}_{\ve{l},\sigma}     \nonumber \\   
& & -\frac{J_kV}{4} \sum_{{\ve{l}},\sigma} \left(  \hat{c}^{\dagger}_{{\ve{l}},0,\sigma}  \hat{d}^{}_{\ve{l},0,\sigma} + h.c. \right) - \lambda  \sum_{{\ve{l}},\sigma} \hat{d}^{\dagger}_{{\ve{l}},\sigma}\hat{d}^{}_{{\ve{l}},\sigma}  \nonumber 
\end{eqnarray}
describing the hybridization of a band of  spinons   with the conduction electron.   Here  the mean field order parameters  $ V = \langle \hat{c}^{\dagger}_{\ve{l},0,\sigma}  \hat{d}^{}_{\ve{l},0,\sigma}  \rangle $ and $\chi =  \langle  \hat{d}^{\dagger}_{\ve{l},\sigma} \hat{d}^{}_{\ve{l}+1,\sigma} \rangle $ have to be determined self-consistently  and  the Lagrange  multiplier $\lambda$ enforces the constraint on average. 
\begin{figure}[htbp]
\centering
\includegraphics[width=\textwidth]{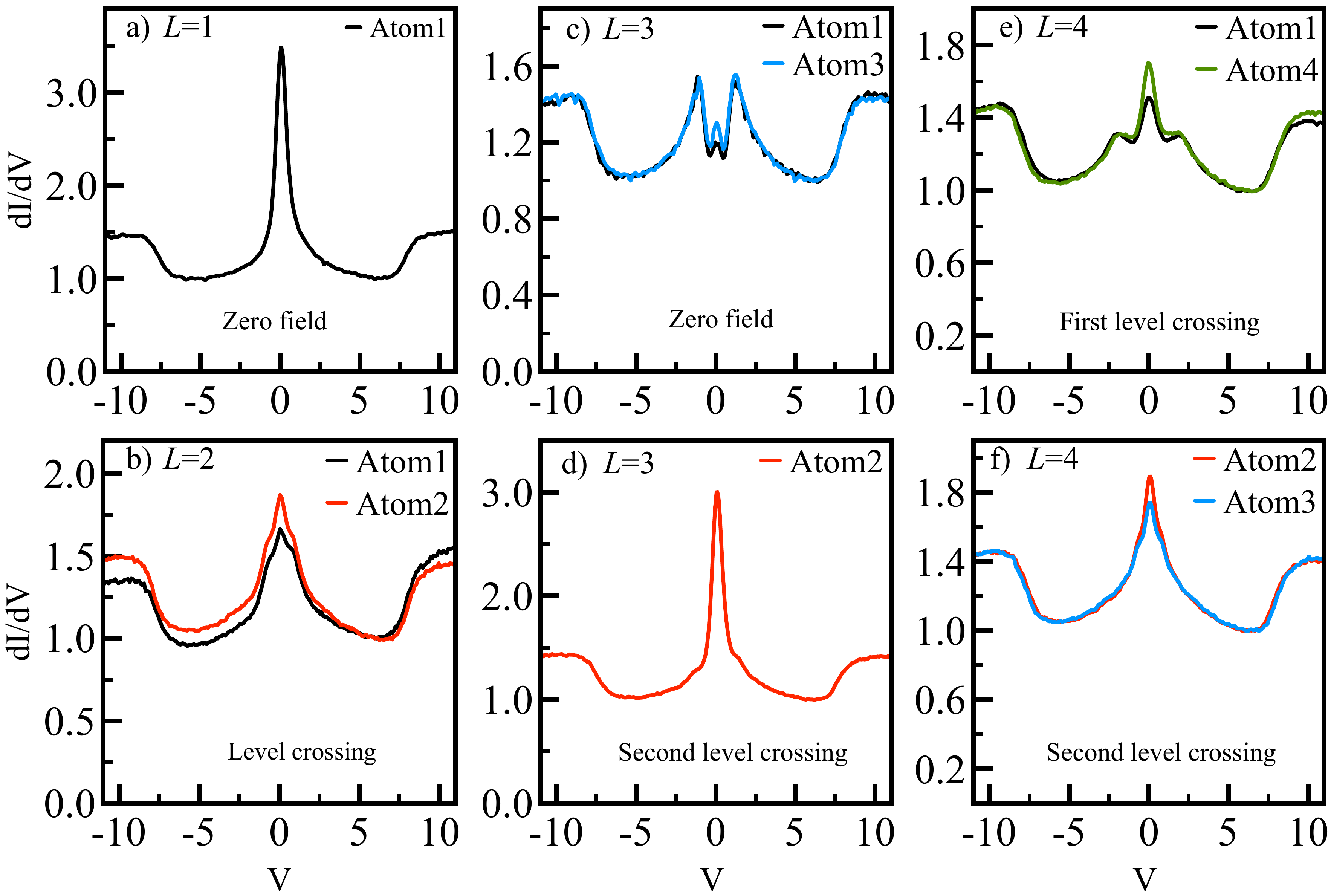}
\caption{Differential conductance (in atomic units)  as a function of voltage measured in STM  experiment of Ref.~[\onlinecite{Toskovic2016}].}
\label{fig:dIbydV_vs_V_STM}
\end{figure}
 \end{document}